\documentclass[12pt,letterpaper]{JHEP3}

\usepackage{amscd,amsmath,amssymb,amsfonts,xspace,mathrsfs}

\numberwithin{equation}{section}

\def\varpi{\mathrm{t}}

\def\sign{{\rm sign}}

\def\Im{\,{\rm Im}\,}

\def\({\left(}
\def\){\right)}
\def\[{\left[}
\def\]{\right]}
\def\hf{{1\over 2}}

\def\<{\left\langle}
\def\>{\right\rangle}

\newcommand{\de}{\mathrm{d}}

\newcommand{\I}{\mathrm{i}}

\newcommand{\cL}{\mathcal{L}}
\newcommand{\cD}{\mathcal{D}}
\newcommand{\ep}{\varepsilon}

\newcommand{\p}{\partial}

\newcommand{\cF}{\mathcal{F}}
\newcommand{\cV}{\mathcal{V}}
\newcommand{\cC}{\mathcal{C}}
\newcommand{\cS}{\mathcal{S}}

\newcommand{\cK}{\mathcal{K}}
\newcommand{\cM}{\mathcal{M}}

\newcommand{\cN}{\mathcal{N}}

\newcommand{\cX}{\mathcal{X}}

\newcommand{\cT}{\mathcal{T}}

\DeclareSymbolFont{AMSa}{U}{msa}{m}{n}
\DeclareSymbolFont{AMSb}{U}{msb}{m}{n}
\DeclareMathSymbol{\fieldR}{\mathalpha}{AMSb}{"52}

\newcommand{\kahler}{{K\"ahler}\xspace}
\newcommand{\hk}{{hyperk\"ahler}\xspace}
\newcommand{\qk}{{quaternion-K\"ahler}\xspace}

\newcommand{\cZ}{\mathcal{Z}}
\newcommand{\cI}{\mathcal{I}}

\newcommand{\cH}{\mathcal{H}}
\newcommand{\cU}{\mathcal{U}}
\newcommand{\cA}{\mathcal{A}}
\newcommand{\cB}{\mathcal{B}}

\newcommand{\nn}{\nonumber}

\newcommand{\IR}{\mathbb{R}}
\newcommand{\IC}{\mathbb{C}}
\newcommand{\IZ}{\mathbb{Z}}

\newcommand{\Nint}{\mathbb{N}}

\newcommand{\talp}{\tilde\alpha}
\newcommand{\teta}{\tilde\eta}

\newcommand{\txi}{\tilde\xi}

\newcommand{\CP}{\IC P^1}

\def\bea{\begin{eqnarray}}
\def\eea{\end{eqnarray}}
\def\be{\begin{equation}}
\def\ee{\end{equation}}
\def\ba{\begin{align}}
\def\ea{\end{align}}
\def\bse{\begin{subequations}}
\def\ese{\end{subequations}}

\def\ba{\bar a}

\def\hX{\hat X}
\def\hY{\hat Y}
\def\hK{\hat K}

\def\ct{\check t}

\def\hcF{\hat\cF}

\def\ci#1{c^{[#1]}}
\def\cij#1{c^{[#1]}}

\newcommand{\Li}{{\rm Li}}

\def\XXint#1#2#3{{\setbox0=\hbox{$#1{#2#3}{\int}$}
\vcenter{\hbox{$#2#3$}}\kern-.5\wd0}}

\def\cij#1{c}
\def\ci#1{c}

\def\Ab{\mathbf{A}}

\def\CY{\mathfrak{Y}}

\def\qli2#1{\Psi_{#1}}
\def\qLif{\mathbf{F}}

\newcommand{\under}[2]{\mathop{#1}\limits_{#2}}

\def\cl0{\tilde c_0}

\def\VT{\cL_\Theta}

\def\chgam{\check\gamma}
\def\chH{\check H}
\def\chcH{\check \cH}

\def\Hil{\mathscr{H}}
\def\Ub{\mathbf{U}}
\def\Ubk{\hat\Ub^{(k)}}

\def\Akhh{\Ab^{(k)}_\hbar}
\def\Akh{\hat\Ab^{(k)}_\hbar}
\def\hAk{\hat\Ab^{(k)}}

\def\il{l}
\def\hST{\hat\cS_T}

\preprint{L2C:15-197 \\
CERN-PH-TH-2015-262\\
arXiv:1511.02892v2}

\title{Theta series, wall-crossing and quantum dilogarithm identities}

\author{Sergei Alexandrov$^{1}$, Boris Pioline$^{2,3}$
\\
$^1$ {\it
Laboratoire Charles Coulomb (L2C), UMR 5221 CNRS-Universit\'e de
Montpellier, F-34095, Montpellier, France}\\

$^2$ {\it CERN PH-TH,
Case C01600, CERN, CH-1211 Geneva 23, Switzerland}\\

$^3$ {\it Laboratoire de Physique Th\'eorique et Hautes
Energies, CNRS UMR 7589, \\
Universit\'e Pierre et Marie Curie,
4 place Jussieu, 75252 Paris cedex 05, France} \\

\vspace*{2mm} {\tt e-mail:
\email{salexand@univ-montp2.fr},
\email{boris.pioline@cern.ch}
}

\vspace*{-3mm}

}

\abstract{
Motivated by mathematical structures which arise in string vacua
and gauge theories with $\cN=2$ supersymmetry, we study the
properties of certain generalized theta series which appear as Fourier
coefficients of functions on a twisted torus. In Calabi-Yau string vacua,
such theta series encode instanton corrections from $k$ Neveu-Schwarz five-branes.
The theta series are determined by vector-valued wave-functions, and in this work
we obtain the transformation of these wave-functions induced by Kontsevich-Soibelman
symplectomorphisms. This effectively provides a quantum version of these transformations,
where the quantization parameter is inversely proportional to the five-brane charge $k$.
Consistency with wall-crossing implies a new five-term relation for
Faddeev's quantum dilogarithm $\Phi_b$ at $b=1$, which we prove.
By allowing the torus to be non-commutative, we obtain a more general five-term
relation valid for arbitrary $b$ and $k$, which may be relevant for the physics
of five-branes at finite chemical potential for angular momentum.}


\dedicated{Mathematics Subject Classification (2010): 13F60, 33B30, 53C26, 81R05, 81T30, 81T60}

\begin{document}

\section{Introduction}

In this work, we study the behavior of certain generalized theta series under a set of contact transformations which arise
naturally in various physical and mathematical contexts, including gauge theories and string vacua with $\cN=2$ supersymmetry,
quaternionic geometry and cluster varieties. Furthermore, we exploit these results to establish new dilogarithm identities.
Before explaining our motivations, let us define the objects of interest and state the mathematical problem that we aim to solve.

\subsection*{Mathematical goal \label{sec_goal}}

Let $(\xi^\Lambda,\txi_\Lambda,\talp)$, with $\Lambda=1,\dots,d$, be coordinates\footnote{Our notations are chosen to match
the literature on hypermultiplet moduli spaces in Calabi-Yau string vacua \cite{Alexandrov:2011va,Alexandrov:2013yva}.
Repeated indices are implicitly summed over, and  indices are often omitted to avoid cluttering.}
on $\IC^{2d+1}$. This space carries an action of the discrete Heisenberg group ${\rm Heis}_{2d+1}$ given
by
\be
T_{\eta^\Lambda,\teta_\Lambda,\kappa}\ :\ (\xi^\Lambda,\txi_\Lambda,\talp)\ \mapsto\
(\xi^\Lambda+\eta^\Lambda,\, \txi_\Lambda+\teta_\Lambda,\, \talp+2\kappa-\teta_\Lambda\xi^\Lambda+\eta^\Lambda\txi_\Lambda-\eta^\Lambda\teta_\Lambda) ,
\label{Heisen}
\ee
where the parameters $\eta^\Lambda,\teta_\Lambda,\kappa$ of the transformation  are all assumed to be integer.
The quotient $\tilde\cT=\IC^{2d+1}/{\rm Heis}_{2d+1}$ is a twisted torus, namely a $\IC^\times$-bundle over
the algebraic torus $\cT=\IC^{2d}/\IZ^{2d}$ parametrized by $(\xi^\Lambda,\txi_\Lambda)$, with fiber parametrized by $e^{\I\pi\talp}$.
In this paper we are interested in functions $H(\xi,\txi,\talp)$ on the twisted torus $\tilde\cT$,
or equivalently functions on $\IC^{2d+1}$ which are invariant under \eqref{Heisen}.

Any such function is a periodic function of the coordinate $\talp$, so can be decomposed in Fourier modes,
\be
H(\xi,\txi,\talp)=
\sum_{k\in\IZ}H_{k}(\xi,\txi) \,e^{-\pi\I k \talp} .
\label{fullfun}
\ee
The zero-mode $H_0(\xi,\txi)$ is a function on the torus $\cT$, which can be further decomposed
in Fourier modes with respect to $\xi^\Lambda$ and $\txi_\Lambda$,
\be
\label{symseries0}
H_0(\xi,\txi) = \sum_{\gamma=(p^\Lambda,q_\Lambda)\in \IZ^{2d}}
h_\gamma \,e^{-2\pi\I(q_\Lambda\xi^\Lambda-p^\Lambda \txi_\Lambda)} .
\ee
In contrast, for $k\neq 0$, $H_{k}(\xi,\txi)$ is a section of the $k$-th power of the Theta line bundle $\VT^k$ over $\cT$,
defined by the quasi-periodicity condition
\be
H_k(\xi+\eta,\txi+\teta)=(-1)^{k\eta^\Lambda\teta_\Lambda}e^{\pi\I k(\eta^\Lambda\txi_\Lambda-\teta_\Lambda\xi^\Lambda)} H_k(\xi,\txi).
\label{thetaline}
\ee
Upon diagonalizing the generators $T_{0,\teta_\Lambda,0}$, this section can
be decomposed into a sum of generalized theta series
\be
H_{k}(\xi,\txi) = \sum_{l^\Lambda\in \frac{\IZ^d}{|k|\IZ^d}}
\sum_{n^\Lambda \in \IZ^d + \frac{l^\Lambda}{k} }
\cH_{k,l^\Lambda}(\xi^\Lambda - n^\Lambda)\,
e^{2\pi\I k n^\Lambda \txi_\Lambda-\pi\I k \xi^\Lambda\txi_\Lambda} .
\label{symseries}
\ee
Note that, whereas the zero-mode $H_0$ is parametrized by a set of constants $h_\gamma$, the other modes
are determined by $|k|^d$-dimensional vectors of functions $\cH_{k,l^\Lambda}(\xi)$,
as a consequence of
the non-commutativity of the translations $T_{\eta^\Lambda,0,0}$ and $T_{0,\teta_\Lambda,0}$ along the torus $\cT$.
The kernels $\cH_{k,l^\Lambda}(\xi)$ are sometimes known as the non-Abelian Fourier coefficients of $H(\xi,\txi,\talp)$.

Crucially for us, the twisted torus $\tilde\cT$ carries a contact structure
given by (the kernel of) the one-form
\be
\label{contactone}
\cX=\de\talp-\xi^\Lambda \de\txi_\Lambda+\txi_\Lambda\de\xi^\Lambda .
\ee
This one-form is invariant under the action \eqref{Heisen} of the Heisenberg group and therefore descends to the quotient $\tilde\cT$.
This paper will focus on the action on functions $H(\xi,\txi,\talp)$ induced by contact transformations on $\tilde\cT$,
i.e.  transformations of $(\xi^\Lambda,\txi_\Lambda,\talp)$ which leave $\cX$ invariant up to rescaling,
and which commute with the Heisenberg group action \eqref{contactone}. In particular, we wish to understand how
these contact transformations act on the Fourier coefficients of $H(\xi,\txi,\talp)$.

A simple class of contact transformations are integer symplectic transformations $\cS$ of the vector\footnote{In order to be
consistent with the Heisenberg action \eqref{Heisen}, the integer symplectic transformation must
be combined with a half-integer shift of $(\xi,\txi)$, see \eqref{sympxi}.
As explained in Appendix \ref{ap-symplectic}, one could restore a homogeneous linear action on $(\xi^\Lambda,\txi_\Lambda)$
at the expense of introducing characteristics in \eqref{Heisen}.}
$(\xi^\Lambda,\txi_\Lambda)$, keeping $\talp$ fixed. Their action on the Fourier coefficients of $H$ is well-known:
while the Abelian Fourier coefficients transform as $h_\gamma\mapsto h_{\cS\cdot \gamma}$ (up to a phase),
the non-Abelian Fourier coefficients $\cH_{k,l^\Lambda}(\xi)$ transform under the metaplectic representation of $Sp(2d,\IZ)$ (see \eqref{metampl}).
In physics terms this means that the kernel $\cH_{k,l^\Lambda}(\xi)$ can be interpreted as a quantum mechanical wave-function.
For example, the symplectic transformation $(\xi^\Lambda,\txi_\Lambda)\mapsto (\txi_\Lambda,-\xi^\Lambda)$ maps $\cH_{k,l^\Lambda}(\xi)$
to its Fourier transform in $\xi$ combined with the discrete Fourier transform in $l^\Lambda$,
which is the standard relation between wave functions in canonically conjugate representations.
This wave function property will be essential in our discussion.

Our main interest will be in a less familiar class of contact transformations $V_\gamma$ on $\tilde\cT$,
obtained by lifting the Kontsevich-Soibelman (KS) symplectomorphisms on $\cT$ \cite{ks}. Their action is best expressed in terms of
the twisted Fourier modes  $\cX_\gamma=\sigma(\gamma)e^{-2\pi\I (q_\Lambda \xi^\Lambda-p^\Lambda\txi_\Lambda)}$,
where $\gamma=(p^\Lambda,q_\Lambda)$ is a vector in the lattice $\IZ^{2d}=\IZ^{d}\oplus \IZ^d$,
equipped with the integer pairing
$\langle \gamma,\gamma'\rangle = q_\Lambda p'^\Lambda - q'_\Lambda p^\Lambda$,
and $\sigma(\gamma)$ is a quadratic refinement of this pairing  (see \eqref{qrf}):
\be
V_\gamma\ : \ \begin{array}{rcl}
\xi^\Lambda&\ \mapsto\ &
\xi'^\Lambda = \xi^\Lambda+\frac{p^\Lambda }{2\pi\I} \, \log\(1-\cX_{\gamma}\), \\
\txi_\Lambda&\ \mapsto\ &
\txi'_\Lambda = \txi_\Lambda+\frac{ q_\Lambda }{2\pi\I}\, \log\(1-\cX_{\gamma}\), \\
\talp\  &\ \mapsto\ &
\talp'=\talp+\frac{1}{2\pi^2}\, L_{\sigma(\gamma)}(\cX_\gamma) .
\end{array}
\label{VKStrans}
\ee
Here, $L_\sigma(z)$ is the Rogers dilogarithm, twisted by $\sigma$ (see \eqref{Rdilog}).
Ignoring the action on the coordinate $\talp$, $V_\gamma$ reduces to the standard KS
symplectomorphism
\be
U_\gamma\ :\ \cX_{\gamma'}\  \mapsto\displaystyle \ \cX_{\gamma'}'=\cX_{\gamma'}\(1-\cX_{\gamma}\)^{\langle\gamma,\gamma'\rangle}.
\label{KStrans}
\ee
Applying $V_\gamma$ to the Fourier decomposition \eqref{fullfun}, one obtains that the generalized theta series
change according to
\be
\label{VKStransH}
H_k(\xi,\txi)\ \mapsto\  H'_k(\xi,\txi) = H_k(\xi', \txi')\, e^{\frac{k}{2\pi\I}\, L_{\sigma(\gamma)}(\cX_\gamma)} .
\ee
Our goal will be to find the action induced by this transformation on the kernels $\cH_{k,l^\Lambda}(\xi)$,
and  verify that this action is consistent with the wave function property.
This will provide a natural quantization $\Ubk_\gamma$ of the symplectomorphism $U_\gamma$
and, furthermore, lead to apparently novel integral identities for the quantum dilogarithm, which turns out to govern the transformation
of the kernels.

\subsection*{Motivation 1: hypermultiplet moduli space in CY vacua}

Our main motivation\footnote{The mathematically minded reader may wish to ignore motivations 1 and 2 and skip to motivation 3.}
for this problem comes from the study of the hypermultiplet moduli space
in Calabi-Yau (CY) compactifications of type II string theory (see \cite{Alexandrov:2011va,Alexandrov:2013yva} for reviews of recent progress).
The \qk (QK) metric on the hypermultiplet moduli space $\cM$ is expected to receive instanton corrections both from Euclidean D-branes and
Neveu-Schwarz (NS) 5-branes wrapping supersymmetric cycles in the CY
threefold $\CY$ \cite{Becker:1995kb}.
Moreover, it must carry an isometric action of  the Heisenberg group
${\rm Heis}_{2d+1}$ with $d=\hf\,b_3(\CY)$ in type IIA string theory (or
$d=b_2(\CY)+1$ in type IIB, respectively), of the monodromy group $\Gamma\subset Sp(2d,\IZ)$
acting on the space $\cS\cK$
of complex structures (or complexified \kahler structures, respectively), and of the modular group
$SL(2,\IZ)$, originating from the S-duality of type IIB strings.
The most efficient way to implement the instanton corrections is to introduce the twistor space $\cZ$,
a $\CP$ fibration over the original manifold $\cM$ endowed with a canonical
complex contact structure. In this framework, the QK metric on $\cM$ is encoded in a set of
complex contact transformations $V^{[ij]}$ between local Darboux coordinate
systems $(\xi^\Lambda,\txi_\Lambda,\talp)$ on an open covering $\{\cU_i\}$ of $\cZ$, such that
the contact structure is locally given by \eqref{contactone}. These complex contact transformations
are conveniently parametrized in terms of holomorphic functions $H^{[ij]}(\xi,\txi,\alpha)$, such that
$V^{[ij]}$ is the exponential of the contact Hamiltonian vector field generated by $H^{[ij]}$ \cite{Alexandrov:2008nk,Alexandrov:2014mfa}.
Thus, to incorporate instanton corrections to the metric on $\cM$, it is sufficient to specify
a set of holomorphic functions and the associated refinement of the open covering of $\cZ$.

The current status of the two types of instanton corrections is quite different.
D-instanton corrections are by now well understood \cite{Alexandrov:2008gh,Alexandrov:2009zh}:
they are implemented by transition functions $H_{\gamma}=\frac{\Omega(\gamma)}{4\pi^2} \,{\rm Li}_2(\cX_\gamma)$ across
certain BPS rays $\ell_\gamma$ on $\CP$, where $\Omega(\gamma)$ are generalized Donaldson-Thomas (DT) invariants counting (with sign)
the number of D-instantons with charge $\gamma$, and ${\rm Li_2}(z)$ is the Euler dilogarithm
\eqref{defLi2}. The corresponding action on the
Darboux coordinates $(\xi^\Lambda,\txi_\Lambda,\talp)$ is precisely the contact transformation
$V_\gamma$ introduced in \eqref{VKStrans}, raised to the power $\Omega(\gamma)$,
which explains the special role played by $V_\gamma$ in our construction.
While the location of the BPS rays $\ell_\gamma$
and the value of $\Omega(\gamma)$  depend on the coordinates in $\cS\cK$, the consistency
of this gluing prescription across the locus where two BPS rays collide is guaranteed by
the Kontsevich-Soibelman wall-crossing formula \cite{ks} and by the dilogarithm identities which follow from it \cite{Alexandrov:2011ac}.
On the other hand, NS5-instanton corrections are still poorly understood.
At the linearized level, they were obtained in \cite{Alexandrov:2010ca} using S-duality of the type IIB formulation.
Although some steps towards extending these results to the non-linear level were taken recently
in \cite{Alexandrov:2014mfa,Alexandrov:2014rca},
a satisfactory twistorial construction is not available yet.
In particular, a general understanding of the compatibility between five-brane instantons
and wall-crossing is still lacking.

In order to make contact with the mathematical problem stated  above, we recall the general
fact that discrete isometries of $\cM$ lift to automorphisms of the complex contact structure of $\cZ$.
This requirement strongly constrains the possible functions $H^{[ij]}$.
In particular, the formal sum of all transition functions must be invariant under the Heisenberg symmetry \eqref{Heisen},
therefore providing an example of the function $H(\xi,\txi,\talp)$ \eqref{fullfun}.
Its zero mode $H_0(\xi,\txi)$ \eqref{symseries0} corresponds to the D-instanton corrections,
while the $k$-th Fourier coefficient $H_k(\xi,\txi)$ with respect to $\talp$ for $k\ne 0$
encodes the contribution from $k$ NS5-brane instantons. As a result, the NS5-brane contributions
can be formally expressed as generalized theta series \eqref{symseries} for a suitable set
of wave functions $\cH_{k,l^\Lambda}(\xi^\Lambda)$.
For $k=1$,  this was demonstrated at the linear level in \cite{Alexandrov:2010ca} using S-duality,
where it was shown that the NS5 wave function $\cH_{1}(\xi^\Lambda)$ is proportional to
(an analytic continuation of) the topological string wave function in the real polarization,
\be
\cH_{1}(\xi)\sim \Psi^{\rm top}_{\IR}(\xi) .
\label{HPsitop}
\ee
This identification relies on interpreting the latter as the partition function of rank one DT invariants \cite{gw-dt},
which determine the contributions from a single D5-brane, and
is in perfect agreement with the wave function interpretation of $\cH_{k,l^\Lambda}(\xi^\Lambda)$.
For $|k|>1$, the five-brane wave-functions $\cH_{k,l^\Lambda}(\xi^\Lambda)$ are expected to
be related to generalized DT invariants for rank $k$ sheaves on $\cX$.

Our result for the transformation property of the wave-functions $\cH_{k,l^\Lambda}$ under
the contact transformations $V_\gamma$ may thus serve as an important clue towards
finding a mutually consistent implementation of D-brane and NS5-brane instantons, however
we shall  not pursue this problem in this work.

\subsection*{Motivation 2: Coulomb branch of $\cN=2$ gauge theories on $\IR^3\times S^1$}

Our second motivation comes from the study of $\cN=2$ gauge theories on $\IR^3\times S^1$ \cite{Seiberg:1996nz}.
The low-energy dynamics of the four-dimensional moduli fields and the holonomies of the electric
and magnetic gauge fields along the circle is described by a non-linear
sigma model. Its target space is a \hk (HK) manifold $\cM'$ of dimension $4d$ where $d$ is
the rank of the gauge group.
When the circle radius $R$ of $S^1$ is very large, $\cM'$ is a torus bundle over
the four-dimensional moduli space $\cS\cK$, with a flat metric along the torus fibers parametrizing the holonomies of the gauge field.
At finite radius, the HK metric on $\cM'$ receives corrections from four-dimensional BPS dyons,
whose Euclidean worldine winds around the circle. Similarly to the instanton contributions to the hypermultiplet metric in string theory,
these corrections are efficiently described
in terms of the twistor space $\cZ'$ of $\cM'$, which is now a complex symplectic manifold with local
Darboux coordinates $(\xi^\Lambda,\txi_\Lambda)$, fibered over $\CP$. As above,  a BPS dyon of charge $\gamma$
induces a discontinuity in the Darboux coordinates  across the BPS ray $\ell_\gamma$,
given by the KS symplectomorphism $U_\gamma$, raised to the power $\Omega(\gamma)$,
where $\Omega(\gamma)$ counts (with sign) the number of four-dimensional BPS states of charge $\gamma$.
Again, the consistency of this prescriptions across walls of marginal stability
is ensured by the KS wall-crossing formula \cite{Gaiotto:2008cd}.

As explained in \cite{Neitzke:2011za}, on top of this \hk structure, $\cM'$ also admits
a canonical hyperholomorphic line bundle $\cL$, controlled by the same data. This line bundle
acquires physical
significance upon placing the $\cN=2$ theory on a Taub-NUT space \cite{Neitzke:2011za,Dey:2014lja}.
By the Ward correspondence,  $\cL$ comes from a holomorphic line bundle $\cL'$ on $\cZ'$. The
gluing conditions for  local sections $\Upsilon(\xi,\txi)$ of $\cL'$ across the BPS ray
$\ell_\gamma$ turn out to coincide precisely with the condition \eqref{VKStransH} for $k=1$,
which allows to identify these sections with our generalized theta series.\footnote{The fact that the same condition
arises both in the context of \hk and \qk geometry is
a consequence of the so-called QK/HK correspondence \cite{Haydys,Alexandrov:2011ac,zbMATH06222565},
which  relates a HK manifold with a rotational isometry endowed with a hyperholomorphic line bundle and
a QK manifold of the same dimension with a quaternionic isometry.}
We expect that our results on the transformation properties of the non-Abelian Fourier coefficients
$\cH_{k,l^\Lambda}(\xi)$ across BPS rays will be useful in understanding the effect of $k$ coincident NUT centers on the \hk sigma model.

We note that in the context of $\cN=2$ gauge theories in four dimensions,
there is a natural quantization arising upon considering line defects and framed BPS states \cite{Gaiotto:2010be}:
the expectation value of a general line defect $L$ with phase $\zeta$ can be expanded as a linear combination
$\sum_\gamma \overline\Omega(L,\gamma,\zeta) \cX_\gamma$ of the exponentiated
Darboux coordinates $\cX_\gamma$, where the framed indices $\overline\Omega(L,\gamma,\zeta)$
count the number of bound states in the presence of $L$
preserving the supersymmetry determined by the phase of $\zeta$.
Upon crossing a BPS ray $\ell_\gamma$ in the $\zeta$-plane, the BPS indices $\overline\Omega(L,\gamma,\zeta)$
jump and the Darboux coordinates experience
the KS symplectomorphism $U_\gamma$, but the expectation value of $L$ stays invariant.
One may refine the index $\overline\Omega(L,\gamma,\zeta)$ into the protected spin character
$\overline\Omega(L,\gamma,\zeta,y)$, keeping track of the angular momentum of the
framed BPS states, at the cost of making the Darboux coordinates non-commutative,
\be
\hat\cX_\gamma\hat\cX_{\gamma'}=q^{\langle\gamma,\gamma'\rangle}\hat\cX_{\gamma'}\hat\cX_\gamma ,
\label{noncomX}
\ee
with $q=y^2$.
Consistency with wall-crossing then requires that the protected spin characters satisfy the so-called motivic wall-crossing formula
\cite{Dimofte:2009bv,Dimofte:2009tm} where the usual dilogarithm is replaced by its quantum version $\Psi_{q}$.
This quantization is in some sense orthogonal to the one induced by the presence of a non-trivial NUT or NS5-brane charge, and
it is natural to ask if a more general
quantization exists which would include both effects at the same time.


\subsection*{Motivation 3: cluster varieties}

Finally, the mathematical problem stated above also arises in the context of cluster varieties introduced
in \cite{2003math11245F,2008InMat.175..223F}. Recall that a cluster variety $\cX$ with
exchange matrix $\cB_{ij}$ of rank $d$ is obtained by gluing together tori $\cT$ with Poisson structure
$\cB_{ij} x_i x_j \partial_{x_i} \partial_{x_j}$ using mutations of $\cB_{ij}$ accompanied by
cluster transformations of the $x_i$ coordinates \cite{2001math4151F,zbMATH05145743}.
A closely related variety $\cA$ is obtained by gluing together tori with symplectic form
$\cB_{ij} \de \log a_i \wedge \de \log a_j$ using suitable transformations of the $a_i$ coordinates.
These structures are particularly useful for deriving functional identities for the
dilogarithm from periodic sequences of mutations \cite{zbMATH06327452}.

To make contact with our mathematical problem, we first recall that cluster transformations
can be decomposed into a `monomial map' $\tau$,
which is an integer change of basis on the torus, and a `birational
automorphism' \cite{2008InMat.175..223F}.
The latter is identical to the KS symplectomorphism $U_\gamma$ (or an integer power thereof,
when the exchange matrix is anti-symmetrizable but not anti-symmetric), provided
the $x_i$-cluster variables are identified with $\cX_\gamma$, up to a change of basis
(see e.g. Appendix B in \cite{Alexandrov:2011ac}).

Second,  one should quantize these cluster varieties. The Poisson variety $\cX$ is quantized by
deformation, i.e.  by replacing the Poisson tori $\cT$ by their non-commutative counterparts $\cT_q$,
such that $\hat x_i \hat x_j=q^{\cB_{ij}} \hat x_j \hat x_i$ \cite{2003math11245F,2008InMat.175..223F}.
This method is well-developed and provides functional identities for the quantum dilogarithm $\Psi_q(x)$.
It is  formally identical to the quantization induced by a non-trivial angular momentum fugacity in the context of $\cN=2$ gauge theories
(cf. \eqref{noncomX}). The symplectic variety $\cA$ is instead usually quantized via  geometric quantization \cite{FGquant}.
The latter is comparatively less developed than the deformation quantization of $\cX$,
but it is the one which is connected with our mathematical problem (and with the previous two physical motivations).
It consists in constructing a vector bundle $\cV$ of rank $|k|^{d}$ over $\cA$,
whose first Chern class is equal to the symplectic form over each torus.
The sections of this vector bundle are in fact our wave functions $\cH_{k,l^\Lambda}(\xi)$,
and their transformation properties under $V_\gamma$, which we establish in this paper,
 provide the gluing conditions which are necessary to specify the bundle over the whole cluster variety $\cA$.

As in deformation quantization, consistency of these transformations with periodic mutation
sequences imply certain functional equations for the non-compact quantum dilogarithm $\Phi_b(x)$.
For the simplest case of the period 5 mutation sequence in the $A_2$ cluster variety, we obtain
an operator identity (see \eqref{pentidentk}) on $L_2(\IR)\otimes \IZ_k$, which we prove by direct computation.
For $k=1$, this relation reduces to the famous quantum pentagon relation
\cite{Chekhov:1999tn,Goncharov:2007,Faddeev:2012zu}\footnote{Here, $\cF$ and $\Phi_b(x)$ are operators on $L_2(\IR)$
acting by Fourier transform and multiplication by $\Phi_b(x)$,
respectively, and $\lambda$ is a computable complex number.}
\be
\(\cF \Phi_b\)^5= \lambda\, \mathbb{I} ,
\label{pentPhi}
\ee
specialized at the value $b=1$. For $k>1$, it involves Fourier transform on both factors
of $L_2(\IR)\otimes \IZ_k$, and multiplication by the quantum dilogarithm
$\Phi_1(x)$, raised to the $k$-th power.

\subsection*{Combining deformation and geometric quantizations}

It is worth noting that, unlike in deformation quantization, the algebra of the Fourier modes
$\cX_\gamma$ remains commutative  in geometric quantization, even though
the quantization parameter $\hbar$ determining the commutation relations of the coordinates $(\xi^\Lambda,\txi_\Lambda)$ is fixed
to $1/k$. This is in accordance with the fact that  the functional identity \eqref{pentidentk} only involves $\Phi_b$ with $b=1$. Furthermore,
 this identity does not seem to follow from the identity \eqref{pentPhi} in the limit where $q=e^{2\pi \I b^2}$ approaches a $k$-th root of identity.
Indeed, the latter limit typically involves the cyclic dilogarithm \cite{Faddeev:1993rs,Cecotti:2010fi,zbMATH06560843},
which does not appear in our set-up. Thus, the two quantization schemes appear in some sense to be orthogonal.

In fact, it is natural to ask if the two schemes can be combined, so that the operator
$\Ubk_\gamma$ arising in geometric quantization is further deformed to an operator which implements
symplectic automorphisms of the non-commutative torus \eqref{noncomX}.
Indeed, physically, there appears to be no obstruction in switching on a non-trivial angular momentum fugacity
in the background of $k$ NS5-branes. In Section 4, we shall implement this idea in the case of
the $A_2$ cluster variety, and produce a deformation \eqref{pentidentkhbar} of the functional identity \eqref{pentidentk},
where the parameter $b$ of the quantum dilogarithm $\Phi_b$ is now arbitrary.
Our derivation of \eqref{pentidentkhbar} relies on a doubling of the non-commutative torus
\eqref{noncomX} analogous to the `modular double' construction of \cite{Faddeev:1995nb},
and could presumably be turned into a mathematical proof along the lines of \cite{Goncharov:2007}.

\subsection*{Relation to complex Chern-Simons theory}

In addition to their relevance for the moduli space of gauge theories and string vacua with 8 supercharges,
and associated wall-crossing phenomena, functional identities for the quantum
dilogarithm also play a central role in the study of quantum Teichm\"uller spaces and Chern-Simons theory
with complex gauge group. Indeed, Chern-Simons theory on a 3-manifold of the form $M=\IR\times \Sigma$,
where $\Sigma$ is a punctured Riemann surface, provides a quantization of
the Teichm\"uller space $\cT(\Sigma)$. Ideal triangulations of $\Sigma$ provide canonical coordinates
on $\cT(\Sigma)$, related by cluster transformations under changes of triangulation, corresponding to
Pachner moves from the point of view of the 3-manifold $M$. Then functional identities
for the quantum dilogarithm ensure invariance under a change of ideal triangulation at the quantum mechanical level.

After completing this paper, we became aware that a similar five-term relation for a quantum dilogarithm
over $L_2(\IR)\otimes \IZ_k$ was obtained in \cite{Andersen:2014aoa} (and, less explicitly,
in \cite{zbMATH06567379}), in the context of complex Chern-Simons theory.
In particular, Eq. (122) in \cite{Andersen:2014aoa} is closely similar to our function \eqref{defAhb}
upon identifying $(N,\theta,x)$ with $(k,b,\sqrt{k} t/b)$, however the limit $q\to 1$ in \cite{Andersen:2014aoa}
corresponds to $\theta=\pm \I$, while in our case it is reached for $b=\pm 1$. It would be interesting to
write down the operatorial identities in \cite{Andersen:2014aoa} in integral form and compare with our construction.
It may also be possible to directly relate, via the 3d-3d correspondence, the partition function of $\cN=2$ theories
of class $S$ on a Taub-NUT space of charge $k$ (as in motivation 2) to complex Chern-Simons
theory on the lens space $L(k,1)$, since $k$-centered Taub-NUT space looks
asymptotically (at least topologically) like $\IR \times L(k,1)$ . We thank the anonymous referee for this suggestion.

\subsection*{Outline}

The organization of the paper is as follows. In section \ref{sec-WC} we derive the action of
the contact transformation $V_\gamma$ on the kernel $\cH_{k,l^\Lambda}(\xi)$ of the theta series,
and prove that it is consistent with the wave function interpretation.
In section \ref{sec-pentagon} we derive the new pentagon relation for the non-compact quantum dilogarithm $\Phi_b$ at $b=1$.
In the final section we construct a further deformation of this identity where the torus becomes
non-commutative and the parameter $b$ is now arbitrary. In Appendix \ref{ap-dilog}, we collect
definitions and useful properties of classical and quantum dilogarithms. In Appendix \ref{ap-symplectic},
we explain how to eliminate the characteristics in the Heisenberg action and in the quadratic refinement
by a variable redefinition. In Appendix \ref{ap-arbch} we prove the identity
\eqref{UXU}, and in Appendix \ref{ap-prop} we collect some useful identities which enter the proof
of Eq. \eqref{indent-k}.

\section{Quantized KS symplectomorphisms}
\label{sec-WC}

Our aim in this section is to derive the action of
the KS symplectomorphism $V_\gamma$ on the kernel $\cH_{k,l^\Lambda}(\xi)$ appearing
in the generalized theta series \eqref{symseries}, and to interpret it as a unitary operator acting on a suitable Hilbert space.
For reasons explained in Appendix \ref{ap-symplectic}, we choose the
quadratic refinement $\sigma(\gamma)$ to be given by
\be
\sigma(\gamma)=(-1)^{q_\Lambda p^\Lambda} .
\label{qrf-sim}
\ee
More general solutions to the quadratic refinement condition \eqref{qrf} can be accommodated by shifting
the coordinates $(\xi^\Lambda,\txi_\Lambda,\talp)$, at the cost of complicating the Heisenberg group action
\eqref{Heisen}.

\subsection{The induced action}
\label{subsec-action}

\subsubsection*{Pure electric case}

Let us first consider the transformation $V_\gamma$ for a charge vector $\gamma=(0,q^\Lambda)$.
Following physics parlance, where $p^\Lambda$ and $q_\Lambda$ correspond to magnetic and electric charges,
respectively, we call such vector `purely electric'.
In this case $\xi^\Lambda$ stays invariant while $\txi_\Lambda$ shifts by a function of $\xi$.
Substituting this transformation into the r.h.s. of \eqref{VKStransH} and using the theta series representation \eqref{symseries},
one arrives at
\be
H_{k}(\xi,\txi) \ \mapsto\
\sum_{l^\Lambda\in \frac{\IZ^d}{|k|\IZ^d}}\sum_{n^\Lambda \in \IZ^d + \frac{l^\Lambda}{k} }
A^{(k)}(\xi,n,q)\, \cH_{k,l}(\xi - n)\, e^{2\pi\I k n^\Lambda \txi_\Lambda-\pi\I k \xi^\Lambda\txi_\Lambda},
\ee
where
\be
A^{(k)}(\xi,n,q)=\(1-e^{-2\pi\I q_\Lambda\xi^\Lambda}\)^{kq_\Lambda n^\Lambda}
\exp\[\frac{k}{2\pi\I}\, L\(e^{-2\pi\I q_\Lambda\xi^\Lambda}\)-\frac{k}{2}\, q_\Lambda\xi^\Lambda\log\(1-e^{-2\pi\I q_\Lambda\xi^\Lambda}\)\].
\label{Axin}
\ee
In view of  the relation \eqref{h1limit}, the exponential is recognized as the $k$-th power of
the Faddeev's quantum dilogarithm function at $\hbar=1$, namely $\qLif_1^k\(-2\pi\I q_\Lambda \xi^\Lambda\)$.\footnote{There are two versions
of the Faddeev's quantum dilogarithm which are commonly used in the literature,
see \eqref{defqLi} and \eqref{relFad}. We denote them by $\Phi_{b}$ and $\qLif_\hbar$, and use either of them  depending
on  convenience.}
Furthermore, due to the periodicity properties \eqref{periodF} of the dilogarithm, the quantity \eqref{Axin} can be rewritten as a function
of only the difference $\xi^\Lambda-n^\Lambda$, which allows to reabsorb it into a redefinition of the kernel.
As a result, the kernel experiences the following transformation
\be
\cH_{k,l^\Lambda}(\xi)\ \mapsto\ \Ab^{(k)}(-q_\Lambda\xi^\Lambda,-q_\Lambda l^\Lambda)
\, \cH_{k,l^\Lambda}(\xi),
\label{trcHk}
\ee
where
\be
\Ab^{(k)}(x,\ell)=\(1-e^{2\pi\I \(x+\ell/k\)}\)^{-\ell}\,\[ \qLif_1\(2\pi\I  \(x+\ell/k\)\)\]^k.
\label{Abkl}
\ee
The periodicity condition $\Ab^{(k)}(x,\ell+k)=\Ab^{(k)}(x,\ell)$ ensures that the transformation
\eqref{trcHk} is independent of the choice of representative of $l^\Lambda$ modulo $|k|\IZ^d$.

\subsubsection*{Generic case}

The transformation of the kernel induced by $V_\gamma$ for a generic charge vector
$\gamma=(p^\Lambda,q_\Lambda)$
can be obtained from the result \eqref{trcHk}, using the fact that $\cH_{k,l^\Lambda}(\xi^\Lambda)$
transform in the metaplectic representation under integer symplectic rotations $Sp(2d,\IZ)$.
Namely, under the three elementary types of such rotations, the kernel transforms as
\cite[(2.39)-(2.41)]{Alexandrov:2010ca}
\begin{subequations}
\bea
\cS_\cA\equiv\begin{pmatrix} \cA^{-T} & 0 \\ 0 & \cA \end{pmatrix}\ &:&\quad
\cH_{k,l^\Lambda}(\xi) \ \mapsto\
\cH_{k,\cA^{-T} l^\Lambda}(\cA^{-T} \xi),
\label{symtrA}
\\
\cS_\cB\equiv \begin{pmatrix} 1 & 0 \\ \cB & 1 \end{pmatrix}\ &:&\quad
\cH_{k,l^\Lambda}(\xi)\ \mapsto\
(-1)^{\cB_{\Lambda\Lambda} l^\Lambda}
e^{\frac{\pi\I}{k}\,\cB_{\Lambda\Sigma}l^\Lambda l^\Sigma- \pi\I k \cB_{\Lambda\Sigma}\xi^\Lambda\xi^\Sigma}
\cH_{k,l^\Lambda}(\xi),
\label{symtrB}
\\
\cS_\cF\equiv\begin{pmatrix} 0 & 1 \\ -1 & 0 \end{pmatrix}\ &:&\quad
\cH_{k,l^\Lambda}(\xi) \ \mapsto\ \(\hcF\cdot \cH_k\)_{l^\Lambda}(\xi),
\label{symtrF}
\eea
\label{metampl}
\end{subequations}
where we defined the Fourier transform operator
\be
\(\hcF\cdot \cH_k\)_{l^\Lambda}(\xi)
=
\sum_{m_\Lambda\in \frac{\IZ^d}{|k|\IZ^d}} e^{-2\pi\I m_\Lambda l^\Lambda/k}
\int \de\txi_\Lambda\, \cH_{k,m_\Lambda}(\txi)\,
e^{2\pi\I k \xi^\Lambda\txi_\Lambda}.
\label{Four-gen}
\ee
Here, the integral runs over the real axis for each $\tilde\xi_\Lambda$.
These transformation rules are easily checked by applying  the coordinate change \eqref{sympxi} on the theta series \eqref{symseries}.
Thus, to obtain the action of $V_\gamma$ on  the kernel $\cH_{k,l^\Lambda}(\xi^\Lambda)$
for an arbitrary charge vector $\gamma=(p^\Lambda,q_\Lambda)$, we need to find an integer
symplectic rotation $\cS$ which maps it to a purely electric vector $(0,q'_\Lambda)$,
apply the transformation $\cS$ on $\cH_{k,l^\Lambda}(\xi^\Lambda)$, then the transformation \eqref{trcHk}
with $q_\Lambda$ replaced by $q'_\Lambda$, and finally $\cS^{-1}$.
The result will be independent of the choice of $\cS$, since the transformations $\cS_{\cA}$ and $\cS_{\cB}$
commute with the operator \eqref{trcHk}.

Unfortunately,  it is difficult to find a combination of integer valued symplectic transformations
$\cS_\cA$, $\cS_\cB$ and $\cS_\cF$ which would be valid for an arbitrary vector $\gamma$.
To circumvent this problem we use the following trick: we extend the torus $\cT=\IC^{2d}/\IZ^{2d}$
to a larger torus $\check{\cT}=\IC^{2d+2}/\IZ^{2d+2}$, parametrized by
$(\xi^I,\txi_I)$ with $I=\flat,1,\dots, d$, and  consider generalized theta series on this auxiliary torus.
We further extend the charge vector $\gamma=(p^\Lambda,q_\Lambda)\in \IZ^{2d}$ to
 $\chgam=(1,p^\Lambda,0,q_\Lambda)\in \IZ^{2d+2}$, and observe that
\be
V_\gamma\cdot e^{-\pi\I k\talp}H_k(\xi^\Lambda,\txi_\Lambda)=\left. V_{\chgam}\cdot e^{-\pi\I k\talp}\chH_k(\xi^I,\txi_I)\right|_{\txi_\flat=0}
\ee
provided the theta series $\chH_k$ is built out the same kernel as the l.h.s.,
namely $\chcH_{k,l^I}(\xi^I)=\cH_{k,l^\Lambda}(\xi^\Lambda)$, with no dependence on $\xi^\flat$ and $l^\flat$.
The virtue of this extension is that it is now easy to find a symplectic transformation mapping $\chgam$
to a purely electric charge vector. Indeed, choosing the symmetric matrix $\cB_{IJ}$ as
\be
\cB_{\flat\flat}=p^\Lambda q_\Lambda,
\qquad
\cB_{\flat\Lambda}=-q_\Lambda,
\qquad
\cB_{\Lambda\Sigma}=0,
\label{matB}
\ee
one has $\cS_\cF\cS_\cB\chgam=\chgam_0$ where $\chgam_0=(0,0,-1,-p^\Lambda)$.
As a result, the transformation of the kernel induced by $V_\gamma$ with a generic charge vector
can be written as\footnote{Note the inverse order of the operators acting on the kernel. This is due to that the map
from operators acting on variables to the operators acting on functions of these variables is an anti-homomorphism. \label{foot-hom}}
\be
V_\gamma\ : \quad
\cH_{k,l^\Lambda}(\xi)\ \mapsto\ \cH'_{k,l^\Lambda}(\xi)=\cS_{\cB}\cdot \hcF\cdot V_{\chgam_0}
\cdot \hcF^{-1}\cdot \cS_\cB^{-1}\cdot \chcH_{k,l^I} .
\label{trcHk-gen}
\ee
Substituting \eqref{Four-gen}, \eqref{symtrB}, \eqref{matB} and \eqref{trcHk} into \eqref{trcHk-gen}, one obtains
an explicit integral form of the transformation. Performing some trivial manipulations, it can be brought to the form
\be
\cH'_{k,l^\Lambda}(\xi)=\sum_{j=0}^{|k|-1}(-1)^{q_\Lambda p^\Lambda j}\, e^{\frac{\pi\I}{k}\,q_\Lambda\( p^\Lambda j^2 +2  l^\Lambda j\)}
\int \de y \,e^{-\pi\I k q_\Lambda\(p^\Lambda y^2+2\xi^\Lambda y\)} \Ab^{(k)}_\cF(y,j) \cH_{k,l^\Lambda+p^\Lambda j}(\xi^\Lambda+p^\Lambda y) ,
\label{integr-trH}
\ee
where $\Ab^{(k)}_\cF(\eta,j)$ is the Fourier transform of the function \eqref{Abkl},
\be
\Ab^{(k)}_\cF(y,j)=\sum_{\ell=0}^{|k|-1}e^{2\pi\I j\ell/k}\int \de x\, e^{-2\pi\I k y x}\, \Ab^{(k)}(x,\ell).
\label{Four-A}
\ee
Note that the dependence on the auxiliary variables $\xi^\flat$ and $l^\flat$ in \eqref{integr-trH} is automatically canceled.
Furthermore, the transformation \eqref{integr-trH} is manifestly independent on the choice of representative
of $l^\Lambda$ modulo $|k| \IZ^{d}$, and for $p^\Lambda=0$ it reduces to the transformation \eqref{trcHk}.
In \eqref{Four-A}, $x$ is integrated along the real axis,
but $\Ab^{(k)}(x,\ell)$ has poles of degree $\ell-kn$ at $x=n-\ell/k$, for any $n\in \IZ$ such that $\ell-kn >0$.
We shall assume these poles are avoided from below.

The integral transformation \eqref{integr-trH} is one of the main results of the paper.
For $k=\pm 1$, the Fourier transform \eqref{Four-A} can be easily evaluated in closed form.
In this case the dependence on $j$ drops out, and the integration contour can be rotated by $\pi/2$
counterclockwise for $k=1$, or clockwise for $k=-1$, so as to reach the imaginary axis.
Due to the choice of contour mentioned in the previous paragraph, no pole is crossed during this process.
Then, using \eqref{FourierF1}, one finds
\be
\Ab^{(\pm 1)}_\cF(y)= \frac{e^{\mp\frac{\pi\I}{12}}\qLif_1^{\mp 1}\(2\pi\I y\)}{1-e^{-2\pi \I y}}.
\label{A-Four}
\ee
For $|k|>1$ we do not know how to evaluate \eqref{Four-A} in closed form.

\subsection{The Hilbert space interpretation}
\label{subsec-Hilbert}

Our next goal is to demonstrate that the transformation of the kernel $\cH_{k,l}(\xi)$
under $V_\gamma$ derived in the previous subsection can be interpreted as the action of a unitary operator
$\Ubk_\gamma$ in a suitable Hilbert space, in line
 with the wave function interpretation of $\cH_{k,l}(\xi)$.
Thus, the operator $\Ubk_\gamma$ provides a quantization of the
KS symplectomorphism $U_\gamma$.


We choose the Hilbert space, which we denote by $\Hil_k=L^2(\IR^d)\otimes \IZ_k^d$,
to be the space of wave functions $\cH_{k,l^\Lambda}(\xi)$
endowed with the following scalar product
\be
\langle \Psi|\Psi'\rangle=
\sum_{l^\Lambda\in \frac{\IZ^d}{|k|\IZ^d}}\int_{\I\IR} \de\xi^\Lambda\, \overline{\Psi_{-l^\Lambda}(\xi)}\,\Psi_{l^\Lambda}'(\xi).
\label{scpr}
\ee
In view of the metaplectic representation \eqref{metampl}, it might seem more natural to consider
the scalar product where $\xi^\Lambda$ is integrated along the real axis and the two wave functions
carry the same index $l^\Lambda$, as opposed to opposite indices $\pm l^\Lambda$.
However, the action of $V_\gamma$ turns out to be unitary provided that $\xi^\Lambda$ takes imaginary values,
and moreover that hermitian conjugation inverts the sign of the discrete variable $l^\Lambda$, as specified in \eqref{scpr}.
With this scalar product, and provided the integration contour
in \eqref{Four-gen} is also rotated to the imaginary axis in a direction such that the integrand stays bounded,
the metaplectic representation \eqref{metampl} is still unitary.

In this space, motivated by our desire that
$\hat\xi^\Lambda$ and $\hat\txi_\Lambda$ be related by the Fourier transform \eqref{Four-gen},
we would like to construct  a representation of the Heisenberg algebra
\be
[\hat\xi^\Lambda,\hat\txi_\Sigma]=\frac{\I}{2\pi k}\,\delta^\Lambda_\Sigma ,
\label{qalg-k}
\ee
where the Planck constant is equated to $1/k$.
In fact, we shall not need the operators $\hat\txi_\Lambda$, but rather their exponentiated
versions $e^{2\pi\I p^\Lambda\hat\txi_\Lambda}$, which in view of \eqref{qalg-k} must satisfy
\be
[e^{2\pi\I p^\Sigma\hat\txi_\Sigma},\hat\xi^\Lambda]=\frac{1}{k}\,p^\Lambda e^{2\pi\I p^\Sigma\hat\txi_\Sigma}.
\ee
To ensure this relation, we take
\be
\hat\xi^\Lambda=\xi^\Lambda\delta_{ll'},
\qquad
e^{2\pi\I p^\Lambda\hat\txi_\Lambda}=e^{\frac{1}{k}\,p^\Lambda\p_{\xi^\Lambda}}\delta_k(l-l'-p),
\label{repr-xi}
\ee
where $\delta_k(n)$ denotes the Kronecker delta function on $\IZ^d/(|k|\IZ^d)$. From
these operators, we construct a quantized version of  the twisted Fourier modes $\cX_\gamma$,
\be
\hat\cX_\gamma=\sigma(\gamma) e^{-2\pi\I q_\Lambda(\hat\xi^\Lambda+l^\Lambda/k)}\,e^{2\pi\I p^\Lambda\hat\txi_\Lambda}.
\label{reprXgam}
\ee
They are hermitian with respect to the scalar product \eqref{scpr}, and satisfy, just like
their classical counterparts, the commutative algebra
\be
\hat\cX_\gamma\hat\cX_{\gamma'}=\hat\cX_{\gamma'}\hat\cX_\gamma =
(-1)^{\langle \gamma,\gamma'\rangle}\, \hat\cX_{\gamma+\gamma'}\, ,
\label{comX}
\ee
despite the fact that $\hat\xi^\Lambda$ and $\hat\txi_\Sigma$ do not commute.
(In particular, the r.h.s. of \eqref{reprXgam} does not depend on the order of the two factors).
Moreover, under the metaplectic representation \eqref{metampl},
\be
\hat\cX_\gamma \mapsto  \hat\cS\, \hat\cX_{\gamma} \, \hat\cS^{-1} = \cX_{\cS\cdot\gamma} .
\ee
For these properties to hold, it is crucial to include the  shift by $l^\Lambda/k$ in the first exponential of \eqref{reprXgam}.
For example, for $d=1$ it is easy to check that Fourier transform turns
the operator
$\hat\cX_{1,0}$ into $\hat\cX_{0,1}$,
namely
\be
\hcF\cdot e^{\p_{\xi}/k}\,\delta_k(l-l'-1)\cdot \hcF^{-1}=e^{-2\pi\I\(\xi+l/k\)}\delta_{ll'} .
\ee

The point of this construction is that the kernel $\cH_{k,l^\Lambda}(\xi)$
of a generalized theta series can now be interpreted
as the overlap $\langle\xi,l|\cH_k\rangle$ between an abstract
state $|\cH_k\rangle$ and a basis of coherent states $|\xi,l\rangle$ which are eigenmodes of
the operator $\hat\xi^\Lambda$. In physics parlance, $\cH_{k,l^\Lambda}(\xi)$
is the wave function of the state $|\cH_k\rangle$ in the $(\xi,l)$-representation.
Moreover, any symplectomorphism of the torus $\cT$ induces a unitary transformation of $\cH_{k,l^\Lambda}(\xi)$,
which can be interpreted as a change of representation of the wave function.
This unitary operator provides a quantization of the corresponding symplectomorphism.
In particular, our construction ensures that integer symplectic rotations are quantized by the metaplectic representation.
Similarly, the action of KS symplectomorphisms on $\cH_{k,l^\Lambda}(\xi)$,
should be described by a unitary operator $\Ubk_{\gamma}$ satisfying the quantized version of  \eqref{KStrans},
\be
\Ubk_\gamma \hat\cX_{\gamma'} \bigl(\Ubk_\gamma\bigr)^{-1}=\hat\cX_{\gamma'}\bigl(1-\hat\cX_{\gamma}\bigr)^{\langle\gamma,\gamma'\rangle} .
\label{UXU}
\ee

Let us now verify this condition for a purely electric charge vector $\gamma=(0,q_\Lambda)$.
In this case, according to the results of the previous subsection, the operator is given by
\be
\Ubk_{(0,q_\Lambda)}=\Ab^{(k)}(-q_\Lambda\xi^\Lambda,-q_\Lambda l^\Lambda)\,\delta_{ll'}.
\label{Uelectric}
\ee
It is indeed unitary as follows from our rules for hermitian conjugation and the explicit form of the function $\Ab^{(k)}$ \eqref{Abkl}.
Substituting \eqref{Uelectric} and \eqref{reprXgam} into \eqref{UXU}, one finds
\bea
\Ubk_{(0,q_\Lambda)} \hat\cX_{\gamma'} \bigl(\Ubk_{(0,q_\Lambda)}\bigr)^{-1}&=&
\sigma(\gamma') e^{-2\pi\I q'_\Lambda(\hat\xi^\Lambda+l^\Lambda/k)}
\Ab^{(k)}(-q_\Lambda\xi^\Lambda,-q_\Lambda l^\Lambda)
\, e^{\frac{1}{k}\,p'^\Lambda\p_{\xi^\Lambda}}\,
\frac{\delta_k(l-l'-p')}{\Ab^{(k)}(-q_\Lambda\xi^\Lambda,-q_\Lambda l'^\Lambda)}
\nn\\
&=&
\bigl(\hat\cX_{\gamma'}\bigr)_{ll'}\,
\frac{\Ab^{(k)}\(-q_\Lambda(\xi^\Lambda-\frac{1}{k}\,p'^\Lambda),
-q_\Lambda (l'^\Lambda+p'^\Lambda)\)}{\Ab^{(k)}(-q_\Lambda\xi^\Lambda,-q_\Lambda l'^\Lambda)}
\nn\\
&=&
\hat\cX_{\gamma'}\(1-e^{-2\pi\I q_\Lambda(\hat\xi^\Lambda+l'^\Lambda/k)}\)^{q_\Lambda p'^\Lambda}.
\label{qKStr}
\eea
This confirms that the operator \eqref{Uelectric}
correctly represents the KS transformation with a pure electric charge.
A slightly more involved computation, outlined in appendix \ref{ap-arbch}, establishes the same result
for the operator $\Ubk_\gamma$ given by the integral transform \eqref{integr-trH}, which implements
the action of the KS symplectomorphism $U_\gamma$ for  a generic charge $\gamma$.

\section{Quantum dilogarithm identities}
\label{sec-pentagon}

As summarized in the introduction, the KS symplectomorphisms considered in the previous section
occur in relation to wall-crossing in supersymmetric field theories or string vacua, and are closely
related to mutations in the context of cluster varieties. In the former context, $U_\gamma$
governs the discontinuity of the Darboux coordinates across a BPS ray $\ell_\gamma$ on the twistor sphere.
Across a wall of marginal stability where two BPS rays $\ell_{\gamma_1}$ and $\ell_{\gamma_2}$ collide,
the KS wall-crossing formula states that the following product of symplectomorphisms must stay constant:
\be
\label{prodKS}
A(\gamma_1,\gamma_2;t)= \prod_{m_1,m_2\geq 0 \atop (m_1,m_2)\neq(0,0)} U_{m_1\gamma_1+m_2\gamma_2}^{\Omega(m_1\gamma_1+m_2\gamma_2;t)} ,
\ee
where the product is ordered so that $m_1/m_2$ increases if $t$ sits on one side of the wall,
or decreases if $t$ sits on the other side. The BPS indices
$\Omega(m_1\gamma_1+m_2\gamma_2;t)$
differ on the two sides, but the effect of the re-ordering of the operators $U_\gamma$ is such that the product \eqref{prodKS}
stays constant. The resulting identity can be rewritten as
\be
\label{prodU}
\prod_{s=1}^{N}U_{\gamma_s}^{\epsilon_s}= \mathbb{I}\, ,
\ee
where $\epsilon_s=\pm 1$ and $\gamma_s$ belong to the set of  charge vectors
$\gamma=m_1\gamma_1+m_2 \gamma_2$ with $\Omega(\gamma;t)\neq 0$ on either side of the wall.
We assume for simplicity that the number $N$ of operators is finite. In the context of string vacua,
smoothness of the metric on the hypermultiplet space requires the apparently stronger condition
that the product \eqref{prodKS}, where the symplectic transformation $U_\gamma$ is replaced by
the contact transformation $V_\gamma$, stays invariant, or equivalently
\be
\label{prodV}
\prod_{s=1}^{N}V_{\gamma_s}^{\epsilon_s}= \mathbb{I}\, .
\ee
In view of the action \eqref{VKStrans} of $V_\gamma$ on the coordinate $\talp$,
this requires a certain functional identity for the Rogers dilogarithm \cite{Alexandrov:2011ac}.
It was shown in \cite{Kashaev:2011se,Alexandrov:2011ac} that this identity in fact follows from the motivic version
of the KS wall-crossing formula in the semi-classical limit $\hbar\to 0$.
The same condition \eqref{prodV} guarantees the existence of a canonical hyperholomorphic line bundle
on the Coulomb branch of $\cN=2$ gauge theories on
$\IR^3\times S^1$ \cite{Neitzke:2011za,Alexandrov:2011ac}. In the context of cluster algebras,
identities of the form \eqref{prodU}, or equivalently \eqref{prodV}, arise from periodic sequences of mutations,
i.e. sequences of mutations which map the exchange matrix $\cB_{ij}$ back to itself.

In the previous section, we have shown that, upon acting on the kernels $\cH_{k,l}(\xi)$ of generalized theta series,
the contact transformations $V_\gamma$ are realized by unitary operators $\Ubk_\gamma$.
Since the map $V_\gamma\mapsto \Ubk_\gamma$ is an anti-homomorphism (see footnote \ref{foot-hom}), for each relation
of the form \eqref{prodV} one obtains an operator identity
\be
\label{prodUhat}
\mathop{\prod_{s=N}}^1\bigl(\Ubk_{\gamma_s}\bigr)^{\epsilon_s}= \mathbb{I}\, ,
\ee
where the order of operators is reversed.
Expressing $\Ubk_\gamma$ as the integral transform \eqref{integr-trH}, for every $k\neq 0$ one arrives at certain integral
identities involving the quantum dilogarithm function $\qLif_1$.
For $k=\pm 1$, the resulting identities turn out to be the specialization at $\hbar=1$ of
the known integral identities satisfied by the quantum dilogarithm $\qLif_{\hbar}$,
which follow from the deformation quantization of cluster varieties \cite{Goncharov:2007,Faddeev:2012zu}.
For $|k|>1$ however, they appear to be genuinely new.

Here we shall consider the example of the five-term identity which arises in the Argyres-Douglas
theory of type $A_2$, or equivalently the $A_2$ cluster variety with
$\cB_{ij}={\scriptsize \begin{pmatrix} 0 & 1\\ -1 & 0 \end{pmatrix}}$:
\be
U_{0,1}^{-1} U_{1,0}^{-1} U_{0,1} U_{1,1} U_{1,0}=\mathbb{I}\,.
\label{pentagonV}
\ee
The quantization of this identity in $\Hil_k=L_2(\IR)\otimes \IZ_k$ implies
\be
\Ubk_{1,0} \Ubk_{1,1}\Ubk_{0,1}\bigl(\Ubk_{1,0}\bigr)^{-1}\bigl(\Ubk_{0,1}\bigr)^{-1}=\mathbb{I}\, .
\label{pentagonhV}
\ee
Below we shall first spell out this operator identity in terms of the quantum dilogarithm $\qLif_1$,
and then provide its direct proof.

Instead of using the integral representation \eqref{integr-trH}, it is convenient to express the operators in \eqref{pentagonhV}
using the symplectic transformations \eqref{metampl} rotating all charges to the electric one $(0,1)$ and
the operator \eqref{Uelectric} specified for this particular charge.
This gives
\be
\begin{split}
\Ubk_{0,1} =&\, \hAk_-,
\\
\Ubk_{1,0} =&\, \hcF\cdot \hAk_+ \cdot \hcF^{-1},
\\
\Ubk_{1,1} =&\,
\hST^{-1}
\cdot \hcF\cdot \hAk_+ \cdot \hcF^{-1}
\cdot \hST,
\end{split}
\label{opk}
\ee
where we introduced two multiplication operators:
\be
\label{defApmS}
\begin{split}
\hAk_{\pm}\ : \quad & \cH_{k,l}(\xi)\ \mapsto\ \Ab^{(k)}(\pm\xi,\pm l)\,  \cH_{k,l}(\xi),
\\
\hST\equiv\hat \cS_{\cB=1}\ : \quad & \cH_{k,l}(\xi)\ \mapsto\ (-1)^{l} e^{-\pi\I( k\xi^2-l^2/k)} \cH_{k,l}(\xi).
\end{split}
\ee
Substituting the expressions \eqref{opk}
and using the properties listed in appendix \ref{ap-prop}, the left-hand side of Eq. \eqref{pentagonhV} becomes
\bea
&&
\hcF\cdot \hAk_+\cdot \hcF^{-1}
\cdot
\hST^{-1}
\cdot \hcF\cdot \hAk_+\cdot \hcF^{-1}\cdot
\hST
\cdot
\hAk_-
\cdot
\hcF\cdot \bigl(\hAk_+\bigr)^{-1}\cdot \hcF^{-1}
\cdot
\bigl(\hAk_-\bigr)^{-1}
\nonumber\\
&=& e^{\frac{7\pi\I k}{12}}
\(\hcF\cdot \hST \cdot \hAk_+ \,\cdot\,\)^5
=e^{-\frac{\pi\I k}{4}}
\(\hcF\cdot \bigl(\hAk_-\bigr)^{-1}\,\cdot\,\)^5.
\label{indent-k}
\eea
Thus, we conclude that the operator on the r.h.s. must be the identity operator.
To write the resulting relation explicitly, we express the function $\Ab^{(k)}$ in terms of $\Phi_1$ using \eqref{relFad},
and change the variables from $\xi$ to $t=-\I\xi$.
Taking into account the sign factors coming from the rotation of the integration contour
in the Fourier transform\footnote{For the integrals appearing in \eqref{indent-k} the integration contours are rotated
clockwise for positive $k$ and counterclockwise for negative $k$, which is the {\it opposite} prescription comparing to the one given for \eqref{Four-A}.
The reason is that all integrands have the function $\Ab^{(k)}$ to the inverse power so that the regions of growth and decay are exchanged.},
we arrive at
\medskip

{\bf \underline{Theorem:}} {\it In $L_2(\IR)\otimes \IZ_k$ spanned by vector valued functions $f_\il(t)$
the following operator identity holds},
\be
\(\sum_{\il'=0}^{|k|-1} e^{-2\pi\I \il\il'/k}\int\limits_{-\infty}^\infty \de t'\, e^{-2\pi\I k tt'}
\,\frac{\[ \Phi_1\(t'-\I \il'/k\)\]^k}{\(1-e^{2\pi\(t'-\I \il'/k\)}\)^{\il'}}\,\,\cdot\, \)^5
=\sign(k)\,\I \, e^{\frac{\pi\I k}{4}}\,\mathbb{I}\, .
\label{pentidentk}
\ee
For $k=1$ this identity reduces to the well-known identity \eqref{pentPhi}
at the special value of the quantum dilogarithm parameter $b=1$.

\subsection*{Proof of the Theorem}

First, we notice that, upon multiplying both sides of the identity by $\hat\cF^{-1}$ from the left and by $\hat\cF$ from the right,
 position of the Fourier transform and the multiplication operator in the brackets are exchanged.
Although this is not necessary for the proof, we find it more convenient to work
with the new operator. Then its kernel is given by
\bea
\cK_{\il_0\il_5}(t_0,t_5)&=& \frac{\[ \Phi_1\(t_0-\I \il_0/k\)\]^k}{\(1-e^{2\pi \(t_0-\I \il_0/k\)}\)^{\il_0}}
\label{ker5}\\
&& \times
\prod_{i=1}^4  \left\{\sum_{\il_i=0}^{|k|-1} \int\limits_{-\infty}^\infty \de t_i \, e^{-2\pi\I \(kt_{i-1}t_i+\frac{\il_{i-1}\il_i}{k}\)}
\,\frac{\[ \Phi_1\(t_i-\I \il_i/k\)\]^k}{\(1-e^{2\pi\(t_i-\I \il_i/k\)}\)^{\il_i}}\right\}e^{-2\pi\I \(k t_4 t_5+\frac{\il_4 \il_5 }{k}\)},
\nonumber
\eea
where we marked the variables $t$ and $\il$ appearing in the integral and the sum of
the $i$th Fourier transform by the index $i$ starting from the left.
To prove \eqref{pentidentk}, we need to show that
\bea
\cK_{\il_0\il_5}(t_0,t_5)
&=& \sign(k)\I e^{\frac{\pi\I k}{4}}\delta_{\il_0\il_5}\delta(t_0-t_5) .
\label{Ker-delta}
\eea

The first step is to shift the integration variables in \eqref{ker5} to absorb the imaginary shift by $\I \il_i/k$.
The integration contour can be shifted back to the real axis since
it does not cross any singularities provided the singularity at $t_i=0$ is avoided by going around it from below or above
depending on whether $\ep=\sign (k)=1$ or $-1$, respectively.
We will do this shift only for the first 3 variables $t_i$, but
we will use the notations
\be
\ct_i=t_i-\I \il_i/k,
\qquad
X_i=e^{2\pi \ct_i}
\label{defXt}
\ee
for all of them. After performing this shift, the kernel \eqref{ker5} becomes
\be
\begin{split}
\cK_{\il_0\il_5}(t_0,t_5)=&\,
\frac{\[ \Phi_1(\ct_0)\]^k}{(1-X_0)^{\il_0}}\sum_{\il_4=0}^{|k|-1}\int\limits_{-\infty}^\infty \de t_4 \,
e^{-2\pi\I \(k t_4 t_5+\frac{ \il_4 \il_5}{k}\)}\frac{\[ \Phi_1(\ct_4)\]^k}{(1-X_4)^{\il_4}}
\\
\times &\,
\prod_{i=1}^3  \left\{ \int\limits_{-\infty}^\infty \de \ct_i  \[e^{-2\pi \I\ct_i\ct_{i-1}}\Phi_1(\ct_i)\]^k
\sum_{\il_i=0}^{|k|-1}\(\frac{ X_{i-1}X_{i+1}}{1-X_i}\)^{\il_i}\right\} X_1^{\il_0}X_3^{\il_4}e^{-2\pi \I k\ct_3\ct_4}.
\end{split}
\label{ker5-shift}
\ee
Next we analyze the integral over $\ct_i$ which is given by the expression in
the curly brackets times $e^{-2\pi \I k\ct_{i+1}\ct_i}$.\footnote{For $i=1$ and $i=3$ there are also factors $X_1^{\il_0}$ and $X_3^{\il_4}$,
respectively. But they are invariant under shifts by $\ct_i\mapsto \ct_i+n\I$, $n\in\IZ$, and therefore do not affect the following considerations.}
The sum over $\il_i$ can be easily evaluated and gives
\be
\begin{split}
&\,
\int\limits_{-\infty}^\infty \de \ct_i  \[e^{-2\pi \I(\ct_{i-1}+\ct_{i+1})\ct_i}\Phi_1(\ct_i)\]^k
\frac{\(\frac{ X_{i-1}X_{i+1}}{1-X_i}\)^{|k|}-1}{\frac{ X_{i-1}X_{i+1}}{1-X_i}-1}
=-\oint_C \de \ct_i  \,
\frac{ \[e^{-2\pi \I(\ct_{i-1}+\ct_{i+1})\ct_i}\Phi_1(\ct_i)\]^k}{\frac{ X_{i-1}X_{i+1}}{1-X_i}-1}\, ,
\end{split}
\label{ker5-int}
\ee
where the closed contour $C$ goes along the real line avoiding $\ct_i=0$ from below or above
depending on $\ep$ and returns back along the line $\Im\ct_i=\ep\I$
avoiding $\ct_i=\ep\I$  from the same side.
Since the dilogarithm function is regular inside this integration contour,
the only singularities of the integrand come from zeros of the denominator. The zeros correspond to solutions of
\be
X_i=1-X_{i-1}X_{i+1}.
\label{recsol}
\ee
Because the strip on the complex plane of $\ct_i$
surrounded by $C$ maps by \eqref{defXt} to the whole complex plane of $X_i$,
one and only one of these solutions lies inside the integration contour.
Thus, the three integrals reduce to the residues at the points related by \eqref{recsol}. The crucial observation is that
this relation defines precisely the periodic sequence \eqref{preiodseq} underlying the pentagon identity.

Let us compute the total contribution to the product of residues coming from the denominator in \eqref{ker5-int}.
To find it, in the $i$th integral we need to express $X_{i-1}$ in terms of $X_0$ and $X_i$ using the recurrence relation \eqref{recsol}.
In particular, one has
\be
X_1=1-X_0 X_2,
\qquad
X_2=\frac{1-X_3}{1-X_0 X_3}.
\ee
Then the total contribution to the three residues from the denominators is given by
\be
\frac{1-X_1}{X_1}\, \frac{1-X_2}{X_2(1-X_0X_3)}\, \frac{(1-X_3)(1-X_0X_3)}{X_3(1-X_4+X_0-2X_0X_3)}=1,
\ee
which is easily checked using \eqref{recsol} and the periodicity of the sequence.
As a result, the expression for the kernel reduces to a single integral
\be
\begin{split}
\cK_{\il_0\il_5}(t_0,t_5)=&\,(-\ep \I)^3\[ \Phi_1(\ct_0)\]^k\sum_{\il_4=0}^{|k|-1}\int\limits_{-\infty}^\infty \de t_4 \,
e^{-2\pi\I \(k t_4 t_5+\frac{\il_4 \il_5 }{k}\)}
\(\frac{ X_1}{1-X_0}\)^{\il_0}\(\frac{X_3}{1-X_4}\)^{\il_4}
\\
&\, \qquad\qquad
\times\prod_{i=1}^4  \[e^{-2\pi \I\ct_i\ct_{i-1}}\Phi_1(\ct_i)\]^k,
\end{split}
\label{evalintker}
\ee
where the variables $X_i$ (and their logarithms $\ct_i$) with $i=1,2,3$ are expressed in terms of $X_0$ and $X_4$ through \eqref{recsol}.

Finally, the use of \eqref{Phib1}, the pentagon identity in the form \eqref{pentagonseq}, and the two relations
$\frac{1-X_0}{X_1}=X_4$ and $\frac{1-X_4}{X_3}=X_0$ following from \eqref{recsol}, allow to show that
\bea
\Phi_1(\ct_0)\prod_{i=1}^4  \[e^{-2\pi \I\ct_i\ct_{i-1}}\Phi_1(\ct_i)\]
&=&\exp\[\frac{\pi\I}{4}
+\frac{\I}{2\pi}\,\log X_0\log X_4\].
\eea
Substituting this result into \eqref{evalintker} and returning back to the original unshifted variables,
we obtain the desired result \eqref{Ker-delta}.
This completes the proof of the Theorem.

\section{Double quantization}
\label{sec-double}

In the previous section, we obtained a quantization of the five-term identity \eqref{pentagonV}
in the Hilbert space $\Hil_k=L^2(\IR)\otimes \IZ_k$, which only involved the quantum dilogarithm $\Phi_b$ at the special value $b=1$.
Moreover, the operators $\Ubk_{\gamma}$, implementing the KS symplectomorphisms \eqref{UXU},
were acting by conjugation on Fourier modes $\hat\cX_{\gamma'}$
which satisfied the commutative algebra \eqref{comX}.
However, it is well-known that  the commutative torus can be deformed to its non-commutative version
\eqref{noncomX}, and that conjugation by the quantum dilogarithm $\Psi_q(\hat\cX_\gamma)$
leads to an automorphism of this non-commutative torus, which reduces to
the KS symplectomorphism $U_\gamma$ in the limit $b\to 1$ \cite{2003math11245F}.
Furthermore, the conjugation by the non-compact quantum
dilogarithm $\Phi_b$ generates automorphisms of two non-commutative tori
$\cT$ and $\cT_\star$ with deformation parameters $q=e^{2\pi i b^2}$ and $q_\star=e^{2\pi\I b^{-2}}$,
respectively, which are mutually commuting \cite{2008InMat.175..223F}.
The product $\cT\times\cT_\star$ is sometimes known as the modular double.
The five-term relation for these automorphisms is guaranteed by a functional identity
for $\Phi_b$ \cite{Goncharov:2007},
which in the special case $b=1$ coincides with our relation \eqref{pentidentk} for $k=1$.
It is thus natural to ask if our construction can also be performed on a genuine non-commutative torus,
providing a functional identity for the $k$-th power of the quantum dilogarithm $\Phi_b$ at arbitrary
values of $b$ and $k\neq 0$.

A physical reason for suspecting that such a double quantization scheme should exist is that the parameter $k$
counts the five-brane charge (or NUT charge, in the field theory setting), while the deformation parameter
$b$ governing the non-commutative torus \eqref{noncomX} is, in the context of line operators and framed BPS states,
conjugate to angular momentum \cite{Gaiotto:2010be}. A priori, there is no physical obstruction to take both $b\neq 1$ and $k\neq 0$.
Indeed, by S-duality the partition function of $k$ NS5-branes in type IIB on a Calabi-Yau threefold $\cX$ is mapped to
the partition function of $k$ D5-branes, with arbitrary D3-D1-D(-1) charge. After compactifying
on a circle and T-dualizing, this is equated to the partition function of $k$ D6-branes with arbitrary D4-D2-D0 charge.
For $k=1$, in the limit where the volume of $\cX$ is very large, the D6-brane can be treated as an infinitely massive source,
or line defect, the D4-brane charge can be removed by a spectral flow, and the partition function then counts framed
BPS states with D2-D0-brane charge. Upon switching on a non-trivial fugacity $y$ for angular momentum,
it becomes a function on the non-commutative torus with $q=y^2$. In particular, it gets conjugated by
$\Psi_q(\hat\cX_\gamma)$ when crossing the BPS wall with charge $\gamma$. For $k>1$,
the D4-brane charge cannot in general be shifted away by spectral flow, but it can be brought to
an element $l^\Lambda\in \IZ^d/|k|\IZ^d$.  Then the corresponding partition function of D2-D0-branes
with $k$ units of D6-brane charge and $l^\Lambda$
units of D4-brane charge, and with non-trivial fugacity for angular momentum,
should be described by some deformation of our wave-functions $\cH_{k,l^\Lambda}$.
In the following, we shall not pursue the physical interpretation of this double quantization,
but we will be content with producing a mathematical
identity generalizing \eqref{pentidentk}, where the $k$-th power of $\Phi_1$ is replaced by $k$ factors
of $\Phi_{b}$, with shifted arguments. Our approach closely follows \cite{Goncharov:2007}
where it was used to prove the standard identity \eqref{pentPhi}.

First, let us introduce a generalization of the function $\Ab^{(k)}$ \eqref{Abkl}
which includes the additional parameter $\hbar$
(or $b=\hbar^{1/2}$)
\be
\begin{split}
\Akhh(t,l)=&\,
\prod\limits_{j=0}^{l-1}\(1-e^{2\pi\(t-\frac{\I l}{k}-\frac{\I(\hbar-1)}{k}\(j+\hf\)\)}\)
\times
\prod\limits_{j=0}^{k-1}
\qLif_{\hbar}\Bigl[2\pi\(t-\tfrac{\I \hbar l}{k}+\tfrac{\I(\hbar-1) }{k}\(\tfrac{k-1}{2}-j\)\)\Bigr] .
\end{split}
\label{defAhb}
\ee
It can be easily checked that this function satisfies the following functional relations,
\be
\begin{split}
\Akhh(t,l)=&\, \Akhh(t,l+k)
\\
=&\,
\left( 1-e^{2\pi\(t-\frac{\I l}{k}-\frac{\I(\hbar-1)}{2k}\)} \right)\, \Akhh(t-\I\hbar/k,l-1)
\\
=&\,
\left( 1-e^{\frac{2\pi}{\hbar}\(t-\frac{\I\hbar l}{k}+\frac{\I(\hbar-1)}{2k}\)} \right)\, \Akhh(t-\I/k,l-1)\, .
\end{split}
\label{propAhb}
\ee

Then, as above, we consider the Hilbert space $\Hil_k=L_2(\IR)\otimes \IZ_k$ whose elements are functions $f_l(t)$
depending on discrete variable $l$ and real $t$. On this space we define two sets of operators
\be
\begin{split}
\hX=&\,  e^{2\pi\(t-\I l/k\)}\delta_{ll'},
\qquad\quad
\hY= e^{-\I\hbar\p_t/k }\delta_k(l-l'-1),
\\
\hX_\star=&\,  e^{2\pi\(t/\hbar-\I l/k\)}\delta_{ll'},
\qquad
\hY_\star= e^{-\I\p_t/k }\delta_k(l-l'-1).
\end{split}
\label{reprXY}
\ee
At $\hbar=1$ these operators are all commutative and reduce to $\hat\cX_{0,1}$ and $\hat\cX_{1,0}$ from section \ref{subsec-Hilbert}.
For any $\hbar$, $\hX$ and $\hY$ still commute with $\hX_\star$ and $\hY_\star$, but between themselves they satisfy
\be
\begin{split}
\hX\hY=&\, e^{2\pi\I\, \frac{\hbar-1}{k}}\hY\hX,
\\
\hX_\star\hY_\star=&\, e^{2\pi\I \frac{1-\hbar}{k\hbar}}\hY_\star\hX_\star.
\end{split}
\ee
This is a particular case of the quantum torus algebra \eqref{noncomX} with the non-commutativity parameters given by
\be
q=e^{2\pi\I\, \frac{\hbar-1}{k}},
\qquad
q_\star=e^{2\pi\I \frac{1-\hbar}{k\hbar}},
\ee
respectively. It is important to note that, unlike in the standard modular double appearing in \cite{2008InMat.175..223F},
the pair $(q,q_\star)$ is {\it not} of the form
$(e^{2\pi\I b^2}, e^{2\pi\I/b^2})$, unless $\hbar=b=1$.
Note also that if we include $\hbar$ in the Fourier transform so that it is defined by
\be
\label{Four-hbar}
\begin{split}
(\hcF_{\hbar}\cdot f )_{l}(t)=& \sum_{l'=0}^{|k|-1} e^{-2\pi\I ll'/k}
\int \de t'\, f_{l'}(t')
e^{-2\pi\I k tt'/\hbar} ,
\end{split}
\ee
then the following relations hold
\be
\begin{split}
\hcF_{\hbar}^{-1} \hX \hcF_{\hbar}=\hY,
&\qquad
\hcF_{\hbar}^{-1} \hY \hcF_{\hbar}=\hX^{-1},
\\
\hcF_{\hbar}^{-1} \hX_\star \hcF_{\hbar}=\hY_\star,
&\qquad
\hcF_{\hbar}^{-1} \hY_\star \hcF_{\hbar}=\hX_\star^{-1} .
\end{split}
\label{FXY}
\ee

We denote by $\Akh=\Akhh(t,l)\delta_{ll'}$  the operator given by the multiplication by the function \eqref{defAhb}.
Using above notations, it can be rewritten as
\be
\begin{split}
\Akh=&\,
\prod_{j=0}^{l-1}\(1-q^{-\hf-j}\hX\)\prod_{j=0}^{k-1}
\qLif_\hbar\(\log\hX+\tfrac{2\pi\I (\hbar-1)}{k}\(\tfrac{k-1}{2}-j -l\)\).
\end{split}
\label{defAhbar}
\ee
The first property in \eqref{propAhb} ensures that $\Akh$ is well defined in $\Hil_k$, whereas
the other two imply the following commutation relations
\be
\begin{split}
\Akh \hY \bigl(\Akh\bigr)^{-1}=&\, \bigl(1-q^{-1/2}\hX\bigr)\,\hY\, ,
\\
\Akh \hY_\star \bigl(\Akh\bigr)^{-1}=&\, \bigl(1-q_\star^{-1/2}\hX_\star\bigr)\,\hY_\star\, .
\end{split}
\label{comXYA}
\ee
We also introduce the operator given by the product of $\Akh$ with the Fourier transform
\be
\hK_\hbar=\hcF_{\hbar}(\Akh\bigr)^{-1}.
\ee
Then, combining \eqref{comXYA} with \eqref{FXY},
one obtains the following commutation relations with the operator $\hK_\hbar$
\be
\begin{split}
\hK_\hbar^{-1}\hX \hK_\hbar=\bigl(1-q^{-1/2}\hX\bigr)\,\hY,
\ &\qquad
\hK_\hbar^{-1}\hY \hK_\hbar=\hX^{-1},
\\
\hK_\hbar^{-1}\hX_\star \hK_\hbar=\bigl(1-q_\star^{-1/2}\hX_\star\bigr)\,\hY_\star,
&\qquad
\hK_\hbar^{-1}\hY_\star \hK_\hbar=\hX_\star^{-1}.
\end{split}
\ee
These relations generate two sequences of operators in $\Hil_k$ which both turns out to be of period 5.
This implies that $\hK_\hbar^5$ commutes with $\hX$, $\hY$, $\hX_\star$ and $\hY_\star$.
In the standard case $k=1$ it was shown in \cite{Goncharov:2007} that this implies that $\hK_\hbar^5$
is proportional to the identity operator. We believe that the same arguments also go through
for arbitrary $k\neq 1$, and imply the following result, which we state as a

\medskip

{\bf \underline{Conjecture:}} {\it In $L_2(\IR)\otimes \IZ_k$ spanned by vector-valued functions $f_l(t)$
the following operator identity holds
\be
\(\sum_{l'=0}^{|k|-1} e^{-2\pi\I ll'/k}\int\limits_{-\infty}^{+\infty} \de t'\, e^{-2\pi\I k tt'/ b^2}\
\frac{\prod\limits_{j=0}^{k-1}\Phi_b\Bigl[b^{-1}t-\frac{\I b l'}{k}+\frac{\I(b-b^{-1}) }{k}\(\frac{k-1}{2}-j\)\Bigr]}
{\prod\limits_{j=0}^{l'-1}\(1-e^{2\pi\(t-\frac{\I l'}{k}-\frac{\I(b^2-1)}{k}\(j+\hf\)\)}\)}
\, \,\cdot\, \)^5= \alpha\, \mathbb{I}\, ,
\label{pentidentkhbar}
\ee
where $\alpha$ is a complex number with unit modulus.}

\medskip

\noindent For $b=1$, \eqref{pentidentkhbar} reduces to \eqref{pentidentk}, providing an alternative derivation of the theorem.

\acknowledgments

It is a pleasure to thank S. Banerjee and D. Persson for valuable discussions.
S.A. thanks the Theory Group at CERN for the kind hospitality throughout the course of this work.

\appendix

\section{Some properties of classical and quantum dilogarithms}
\label{ap-dilog}

\subsection{Euler and Rogers dilogarithms}

The classical Euler dilogarithm is defined for $|z|<1$ by
\be
\label{defLi2}
 {\rm Li}_2(z) = \sum_{n=1}^\infty \frac{z^n}{n^2}  .
\ee
By analytic continuation, it defines a multi-valued function on $\IC$ with
a logarithmic branch cut from $1$ to $\infty$. For $|z|<1$ and $|1-z|<1$, the Rogers dilogarithm is defined by
\be
L(z) = {\rm Li}_2(z)+\frac12\, \log z \log(1-z)\, .
\ee
By analytic continuation, it defines a multi-valued function on $\IC$ with logarithmic branch cuts
from $-\infty$ to $0$ and from $1$ to $\infty$.
Some of the special values of $L(z)$ are
\be
L(0)=0,
\qquad
L(1/2)=\frac{\pi^2}{12} ,
\qquad
L(1) = \frac{\pi^2}{6} .
\ee
The Rogers dilogarithm satisfies the functional relations, valid when all arguments of $L(z)$ satisfy $|z|<1$ and $|1-z|<1$,
\be
L(z)+L(1-z)=L(1) ,
\label{funrelRdil}
\ee
\be
L(x)+L(y)=L(xy)+L\(\frac{x(1-y)}{1-xy}\)+L\(\frac{y(1-x)}{1-xy}\).
\label{pentagon}
\ee
The  five-term identity \eqref{pentagon} can also be rewritten in the following form
\be
\sum_{i=0}^4 L(x_i)=3L(1)=\frac{\pi^2}{2},
\label{pentagonseq}
\ee
where the variables $x_i$ satisfy the recurrence relation
\be
x_i=1-x_{i-1}x_{i+1}
\label{preiodseq}
\ee
which gives a periodic sequence with period 5: $x_{i+5}=x_i$. Explicitly, the elements of the sequence read
\be
x_2=\frac{1-x_1}{x_0},
\qquad
x_3=\frac{x_0+x_1-1}{x_0x_1},
\qquad
x_4=\frac{1-x_0}{x_1},
\qquad
x_5=x_0.
\label{explicitseq}
\ee
To get \eqref{pentagonseq} from \eqref{pentagon}, one should identify $x_0=x$, $x_1=1-xy$ and use
the relation \eqref{funrelRdil} for 3 terms.

The contact transformation \eqref{VKStrans} obtained by lifting the KS symplectomorphism $U_\gamma$ involves
a close cousin of the Rogers dilogarithm which depends on an additional complex parameter $\epsilon$,
\be
L_\epsilon(z)=\Li_2(z)+\hf\, \log(\epsilon^{-1}z)\log(1-z)\, .
\label{Rdilog}
\ee
Its properties can be easily derived from the properties of the standard Rogers dilogarithm.

\subsection{Non-compact quantum dilogarithm}

An important version of the quantum dilogarithm, which is relevant for the theta series studied in this paper,
was introduced by Faddeev in \cite{Faddeev:1995nb}. It is defined in one of the two equivalent ways as follows
\be
\begin{split}
\Phi_b(x) = &\,
\prod_{k=0}^\infty\frac{1+e^{2\I\pi b^2(k+1/2)}e^{2\pi b x}}{1+e^{-2\pi \I b^{-2}(k+1/2)}e^{2\pi b^{-1} x}}
=
\exp\left( \frac14 \int_{\IR+\I 0^+}
\frac{e^{-2\I x t}}{\sinh (b t)\, \sinh (b^{-1}t)} \,\frac{\de t}{t}\right).
\end{split}
\label{defqLi}
\ee
This function has the following properties (see e.g. \cite{Faddeev:2000if}):
\begin{itemize}
\item
$\Phi_b(x)$ is meromorphic in $x$ with
\be
{\rm poles:}\ c_b+\I b \Nint+\I b^{-1}\Nint,
\qquad
{\rm zeros:}\ -c_b-\I b \Nint-\I b^{-1}\Nint,
\ee
where $c_b=\frac{\I}{2}\,(b+b^{-1})$.

\item
Conjugation properties:
\be
\Phi_b(x)\Phi_b(-x)=\Phi_b(0)^2 e^{\pi\I x^2},
\qquad
\Phi_b(0)
=e^{\frac{\pi\I}{24}\, (b^2+b^{-2})},
\ee
\be
(1-|b|)\Im b=0 \ \Rightarrow \ \overline{\Phi_b(x)}=1/\Phi_b(\bar x).
\ee

\item
Quasi-periodicity:
\be
\Phi_b(x-\I b^{\pm 1}/2)=\(1+e^{2\pi b^{\pm 1}x}\)\Phi_b(x+\I b^{\pm 1}/2).
\label{periodPhi}
\ee

\item
Five-term relation:
\be
\Phi_b(q)\, \Phi_b(p) = \Phi_b(p)\,\Phi_b(p+q)\,\Phi_b(q),
\ee
where $p$ and $q$ are non-commutative variables satisfying $[p,q]=1/(2\pi\I)$.

\item
Fourier transform:
\be
\begin{split}
\int_{\IR}\Phi_b(x)e^{2\pi\I w x}\de x=&\, \frac{\zeta_o^{-1} e^{2\pi\I w c_b}}{\Phi_b(-w-c_b)}
=\zeta_o\, e^{-\pi\I w^2}\Phi_b(w+c_b),
\\
\int_{\IR}\(\Phi_b(x)\)^{-1} e^{2\pi\I w x}\de x=&\, \zeta_o\, e^{-2\pi\I w c_b}\Phi_b(w+c_b)
=\frac{\zeta_o^{-1} e^{\pi\I w^2}}{\Phi_b(-w-c_b)},
\end{split}
\label{FourierPhi}
\ee
where $\zeta_o=e^{\frac{\pi\I}{12}\,(1-4c_b^2)}=e^{\frac{\pi\I}{12}\(3+b^2+b^{-2}\)}$.

\item
Asymptotics:
\be
\Phi_b(x)\under{\sim}{|x|\to \infty} \left\{\begin{array}{ccc}
1 & {\rm \ if\ }&|\arg(x)|>\frac{\pi}{2}+\arg b,
\\
\Phi_b(0)^2 e^{\pi\I x^2} & {\rm \ if\ }& |\arg(x)|<\frac{\pi}{2}-\arg b,
\\
C(-b^{-2})/\Theta(\I b^{-1}x, -b^{-2}) & {\rm \ if\ }& |\arg(x-\pi/2)|<\arg b,
\\
\Theta(\I b x, b^2)/C(b^2) & {\rm \ if\ }& |\arg(x+\pi/2)|<\arg b,
\end{array}\right.
\label{asymPhi}
\ee
where
\be
C(u)=\prod_{k=1}^\infty\(1-e^{2\pi\I k u}\),
\qquad
\Theta(z,\tau)=\sum_{n\in \IZ}e^{\pi\I\tau n^2+2\pi\I n z}.
\ee

\item
Classical limit:
\be
\label{claslimit}
\Phi_b(x) \under{\sim}{b\to 0} \exp\left(\frac{{\rm Li}_2(-e^{2\pi b x})}{2\pi\I b^2}\right)  .
\ee

\item
Value at $b=1$:
\be
\Phi_1(x)=\exp\left[\frac{\I}{2\pi}\(\Li_2(e^{2\pi x})+2\pi x \log\(1-e^{2\pi x}\) \)\right].
\label{Phib1}
\ee

\end{itemize}
The last property can be established from \eqref{defqLi} by closing the integration contour
in the upper half plane. In particular, it makes it manifest that $|\Phi_1(x)|=1$ for $x$ real,
and that $\Phi_1(x)$ has a pole of degree $n$ at $x=\I n$ and a zero of degree
$n$ at $x=-\I n$, for any $n=1,2,\dots$.

Sometimes it is more convenient to work in terms of a redefined version of the Faddeev's dilogarithm which is given by
\be
\qLif_{b^2}(z)=\frac{1}{\Phi_b(z/(2\pi b))}.
\label{relFad}
\ee
In view of this definition, it is convenient to introduce the quantization parameter $\hbar=b^2$.
In terms of this function, the periodicity properties \eqref{periodPhi} become
\be
\begin{split}
\qLif_\hbar(z+2\pi\I \hbar) =&\, (1+e^{\pi\I\hbar} \, e^z)\, \qLif_\hbar(z) ,
\\
\qLif_\hbar(z+2\pi\I ) =&\, (1+e^{\pi\I/\hbar} \, e^{z/\hbar})\, \qLif_\hbar(z),
\end{split}
\label{periodF}
\ee
while the relation \eqref{Phib1} becomes
\be
\label{h1limit}
\qLif_1(z)=
 \exp\left[\frac{1}{2\pi\I}\left(
{\rm Li}_2(e^z)+z \log(1-e^z) \right)\right].
\ee
Rewriting the Fourier integrals \eqref{FourierPhi} in terms of this function and specializing them to the case $\hbar=1$,
one finds
\be
\begin{split}
\int_{\IR}\frac{e^{2\pi\I w x}}{\qLif_1(2\pi x)}\,\de x=&\, \frac{e^{\frac{7\pi\I}{12}}\qLif_1(-2\pi w)}{1-e^{2\pi w}}
= \frac{e^{\frac{5\pi\I}{12}-\pi\I w^2}}{\(1-e^{2\pi w}\)\qLif_1(2\pi w)},
\\
\int_{\IR}\qLif_1(2\pi x) e^{2\pi\I w x}\de x=&\,\frac{e^{-\frac{7\pi\I}{12}}}{\(1-e^{-2\pi w}\)\qLif_1(2\pi w)}
={e^{-\frac{5\pi\I}{12}+\pi\I w^2}}\frac{\qLif_1(-2\pi w)}{1-e^{-2\pi w}},
\end{split}
\label{FourierF1}
\ee
where we took into account that $\zeta_o=e^{\frac{5\pi\I}{12}}$
and
\be
\qLif_1(z)\qLif_1(-z)=e^{-\frac{\pi\I}{6}} e^{-\frac{\I z^2}{4\pi}}.
\label{Fbinv}
\ee

\section{Characteristics and symplectic transformations}
\label{ap-symplectic}

The Heisenberg action \eqref{Heisen} is a special case of the more general action
\be
\label{Heisengen}
 (\xi^\Lambda,\txi_\Lambda,\talp)\ \mapsto\
(\xi^\Lambda+\eta^\Lambda,\, \txi_\Lambda+\teta_\Lambda,\,
\talp+2\kappa-\teta_\Lambda\xi^\Lambda+\eta^\Lambda\txi_\Lambda-\eta^\Lambda\teta_\Lambda + \teta_\Lambda \theta^\Lambda-\eta^\Lambda \phi_\Lambda)
\ee
parametrized by  $(\theta^\Lambda,\phi_\Lambda)\in \IR^{2d}/\IZ^{2d}$, known as characteristics (see e.g.  \cite{Alexandrov:2010np}).
The main advantage of introducing these additional parameters is that the standard
linear action of the symplectic group on the vector $(\xi^\Lambda,\txi_\Lambda)$
is compatible with the Heisenberg action \eqref{Heisengen}, provided the characteristics
pick up a half-integer shift,
\be
\label{sympchar}
\begin{pmatrix} \xi^\Lambda \\ \txi_\Lambda \end{pmatrix}
\ \mapsto\
\cS \cdot \begin{pmatrix} \xi^\Lambda \\ \txi_\Lambda \end{pmatrix} ,
\qquad
\talp \mapsto \talp,
\qquad
\begin{pmatrix} \theta^\Lambda \\ \phi_\Lambda \end{pmatrix}
\ \mapsto\
\cS
\cdot \[
\begin{pmatrix} \theta^\Lambda \\ \phi_\Lambda  \end{pmatrix}
-\frac12
\begin{pmatrix} (\cA^T\cC)_d \\ (\cD^T\cB)_d  \end{pmatrix}
\],
\ee
where $\cS={\scriptsize \begin{pmatrix}\cD & \cC \\ \cB & \cA \end{pmatrix}} \in Sp(2d,\IZ)$ and
$(A)_d$ denotes the diagonal of a matrix $A$.
The change of variables
\be
\begin{split}
\xi^\Lambda\ \mapsto\  \xi^\Lambda+\theta^\Lambda,
\qquad
\txi_\Lambda\ \mapsto\ \txi_\Lambda+\phi_\Lambda,
\qquad
\talp\ \mapsto\ \talp+\theta^\Lambda\txi_\Lambda-\phi_\Lambda\xi^\Lambda ,
\end{split}
\label{redef-ch}
\ee
removes the characteristics  from the action \eqref{Heisengen}, so there is no loss of
generality in setting $(\theta^\Lambda,\phi_\Lambda)=0$, as we have done in \eqref{Heisen}.
The price to pay, however, is that the new coordinates $(\xi^\Lambda,\txi_\Lambda,\talp)$
no longer transform homogeneously under symplectic transformations,
rather \cite{Alexandrov:2014rca}
\be
\label{sympxi}
\begin{split}
\begin{pmatrix} \xi^\Lambda \\ \txi_\Lambda \end{pmatrix}
\ \mapsto\ &\,
\cS
\cdot \[
\begin{pmatrix} \xi^\Lambda \\ \txi_\Lambda  \end{pmatrix}
+\frac12
\begin{pmatrix} (\cA^T\cC)_d \\ (\cD^T\cB)_d  \end{pmatrix}
\],
\\
\talp \ \mapsto\ &\,  \talp +\hf\((\cA^T \cC)^{\Lambda\Lambda}\txi_\Lambda-(\cD^T\cB)_{\Lambda\Lambda}\xi^\Lambda\).
\end{split}
\ee
It is this action that is to be used in section \ref{sec-WC}, when checking the symplectic
action on the kernels of theta series.

A similar freedom arises in the quadratic refinement $\sigma(\gamma)$, which is defined by the condition
\be
\sigma(\gamma)\, \sigma(\gamma') = (-1)^{\langle\gamma,\gamma'\rangle} \sigma(\gamma+\gamma').
\label{qrf}
\ee
The most general solution is also parametrized by characteristics $(\theta'^\Lambda,\phi'_\Lambda)$:
\be
\sigma(\gamma)=e^{-\pi\I p^\Lambda q_\Lambda + 2\pi\I (q_\Lambda \theta'^\Lambda - p^\Lambda \phi'_\Lambda)} .
\label{qrf-ch}
\ee
In particular, one may check that the KS-symplectomorphism $V_\gamma$ in \eqref{VKStrans}
commutes with the Heisenberg action \eqref{Heisengen} for arbitrary choices of characteristics.
As above, the change of variables \eqref{redef-ch} allows to set $(\theta'^\Lambda,\phi'_\Lambda)$ to zero.
However, there is no way to set both $(\theta^\Lambda,\phi_\Lambda)$ and $(\theta'^\Lambda,\phi'_\Lambda)$ to zero,
unless they happen to coincide. Interestingly, it was shown that in the string theory context, this condition follows from
S-duality \cite{Alexandrov:2014rca}.
With this motivation in hindsight, in this work we set all characteristics to zero,
keeping in mind that the coordinates $(\xi^\Lambda,\txi_\Lambda,\talp)$
transform non-homogeneously under symplectic rotations as in \eqref{sympxi}.

\section{Action of KS symplectomorphism with arbitrary charge}
\label{ap-arbch}

In this appendix we verify that the unitary operator $\Ubk_\gamma$
defined by the integral transformation \eqref{integr-trH} generates the standard KS action on the operators $\hat\cX_{\gamma'}$,
i.e. satisfies the condition \eqref{UXU}.
Explicitly one obtains
\bea
\Ubk_\gamma \hat\cX_{\gamma'} \bigl(\Ubk_\gamma\bigr)^{-1}\cdot && \cH_{k,l^\Lambda}=
\sum_{j=0}^{|k|-1}(-1)^{q_\Lambda p^\Lambda j}\, e^{\frac{\pi\I}{k}\,q_\Lambda\( p^\Lambda j^2 +2  l^\Lambda j\)}
\int \de y \,e^{-\pi\I k q_\Lambda\(p^\Lambda y^2+2\xi^\Lambda y\)} \Ab^{(k)}_\cF(y,j)
\nn\\
&&\times
\sigma(\gamma') e^{-2\pi\I q'_\Lambda\(\xi^\Lambda+\frac{l^\Lambda}{k}+p^\Lambda\(y+\frac{j}{k}\)\)}\,
e^{\frac{1}{k}\,p^\Lambda\p_{\xi^\Lambda}}
\nn\\
&&\times
\sum_{j'=0}^{|k|-1}(-1)^{q_\Lambda p^\Lambda j'}\, e^{\frac{\pi\I}{k}\,q_\Lambda\( p^\Lambda (j')^2 +2  (l^\Lambda-p'^\Lambda+p^\Lambda j)j'\)}
\!\int \de y' \,e^{-\pi\I k q_\Lambda\(p^\Lambda (y')^2+2(\xi^\Lambda+p^\Lambda y)y'\)}
\nn\\
&&\times
\Ab^{-(k)}_\cF(y',j')
\cH_{k,l^\Lambda-p'^\Lambda+p^\Lambda (j+j')}\(\xi^\Lambda+p^\Lambda (y+y')\),
\eea
where $\Ab^{-(k)}_\cF$ is the Fourier transform of $\(\Ab^{(k)}\)^{-1}$.
Redefining
$y\mapsto y-y'$, $j\mapsto j-j'$ and bringing the derivative operator to the right, one finds
\be
\begin{split}
&\,
\sum_{j=0}^{|k|-1}(-1)^{q_\Lambda p^\Lambda j}\, e^{\frac{\pi\I}{k}\,q_\Lambda\( p^\Lambda j^2 +2  l^\Lambda j\)}
\int \de y \,e^{-\pi\I k q_\Lambda\(p^\Lambda y^2+2\xi^\Lambda y\)}
\cI^{(k)}(y,j)
\\
\times &\,
\sigma(\gamma') e^{-2\pi\I q'_\Lambda\(\xi^\Lambda+\frac{l^\Lambda}{k}+p^\Lambda\(y+\frac{j}{k}\)\)}\,
e^{\frac{1}{k}\,p^\Lambda\p_{\xi^\Lambda}}
\cH_{k,l^\Lambda+p^\Lambda j}\(\xi^\Lambda+p^\Lambda y\),
\end{split}
\label{UXUgamma}
\ee
where we introduced
\be
\cI^{(k)}(y,j)=\sum_{j'=0}^{|k|-1}
\int \de y' \, e^{-2\pi\I \langle\gamma,\gamma'\rangle\(y'+\frac{j'}{k}\)}
\Ab^{(k)}_\cF(y-y',j-j')
\Ab^{-(k)}_\cF(y',j').
\ee
This convolution of the two Fourier transforms can be simply evaluated. Substituting the definition \eqref{Four-A}
and using \eqref{Abkl}, one arrives at a very simple result
\be
\cI^{(k)}(y,j)=\sum_{\ell=0}^{|k|-1}e^{2\pi\I j\ell/k}\int \de x\, e^{-2\pi\I k y x}\(1-e^{2\pi\I\(x+\ell/k\)}\)^{\langle\gamma,\gamma'\rangle}.
\ee
Comparing the resulting expression \eqref{UXUgamma} with \eqref{integr-trH}, one concludes that the only difference is that
$\Ab^{(k)}(x,\ell)$ is replaced here by $\(1-e^{2\pi\I\(x+\ell/k\)}\)^{\langle\gamma,\gamma'\rangle}$ and $\cH_{k,l^\Lambda}(\xi)$
is acted upon by $\hat\cX_{\gamma'}$. Therefore, this expression can be rewritten as (cf. \eqref{trcHk-gen})
\be
\cS_{\cB}\cdot \hcF\cdot \bigl(1-\hat\cX_{\chgam_0}\bigr)^{\langle\gamma,\gamma'\rangle}
\cdot \hcF^{-1}\cdot \cS_\cB^{-1}\cdot \hat\cX_{\gamma'}\cdot\chcH_{k,l^I}
=\bigl(1-\hat\cX_{\gamma}\bigr)^{\langle\gamma,\gamma'\rangle}\hat\cX_{\gamma'}\cdot\cH_{k,l^\Lambda},
\ee
where we used notations from section \ref{subsec-action}.
Due to  \eqref{comX}, this is the same as the r.h.s. of \eqref{UXU},
which proves the desired property.

\section{Some useful properties}
\label{ap-prop}

In section \ref{sec-pentagon} we need some properties of the Fourier transform \eqref{Four-gen} (for $d=1$)
and the operators $\hST$ and $\hAk_{\pm}$ \eqref{defApmS}
acting on $L_2(\IR)\otimes \IZ_k$.

First, it is immediate to check that for any $f_l\in L_2(\IR)\otimes \IZ_k$, one has
\be
\begin{split}
\hcF^{-1}\cdot f_l(\xi)\cdot\hcF =&\,\hcF\cdot  f_{-l}(-\xi)\cdot \hcF^{-1},
\\
\hcF^{-1}\cdot f_l(\xi)\cdot\hcF^{-1} =&\,\hcF\cdot  f_{-l}(-\xi)\cdot \hcF.
\end{split}
\label{Four-inv}
\ee
Second, using the following identity
for shifted quadratic Gauss sums (equivalent to Eq. (5.5) in \cite{Glasser:2014})
\be
\sum_{j=0}^{|k|-1} e^{-\frac{\pi\I}{k}\(j-\frac{k}{2}\)^2}=\sqrt{k}\, e^{-\frac{\pi\I }{4}} ,
\ee
one can show that
\be
\hcF^{-1} \cdot \hST^{-1} \cdot \hcF
=e^{\frac{\pi\I k}{4}} \, \hST\cdot\hcF\cdot \hST\, .
\label{Four-Gaus}
\ee
Finally, using \eqref{Fbinv}, one can establish the following relation between all three operators in \eqref{defApmS}
\be
\begin{split}
\bigl(\hAk_{+}\bigr)^{-1}
=&\, e^{\frac{\pi\I k}{6}}\,\hST\, \hAk_{-}.
\end{split}
\label{propA}
\ee

\providecommand{\href}[2]{#2}\begingroup\raggedright\endgroup


\end{document}